\documentclass[10pt]{article}

\pdfoutput=1 

\usepackage{jheppub} 

\usepackage[T1]{fontenc} 

\usepackage{amssymb,amsthm,cancel,hyperref,graphicx,xcolor}
\usepackage{mathrsfs,wasysym}
\usepackage{booktabs}
\usepackage{tikz}
\usetikzlibrary{calc}
\usepackage{placeins}

\newcommand{\CHbar}{\overline{\text{CH}}}
\newcommand{\sss}{\scriptscriptstyle}

\newcommand{\Sud}{\mathcal{S}}
\newcommand{\as}{\alpha_s}

\newcommand{\Ord}{\mathcal{O}}
\newcommand{\Lum}{\mathscr{L}}

\newcommand{\mh}{m_{\sss\rm H}}
\newcommand{\mt}{m_{\sss\rm t}}

\newcommand{\muf}{\mu_{\sss\rm F}}
\newcommand{\mur}{\mu_{\sss\rm R}}

\let\originalleft\left
\let\originalright\right
\renewcommand{\left}{\mathopen{}\mathclose\bgroup\originalleft}
\renewcommand{\right}{\aftergroup\egroup\originalright}

\def\beq{\begin{equation}}  
\def\eeq{\end{equation}}
\def\({\left(}
\def\){\right)}
\def\[{\left[}
\def\]{\right]}

\title{\boldmath On the Higgs cross section at N$^3$LO+N$^3$LL and its uncertainty}

\author[a]{Marco~Bonvini,}
\author[b]{Simone~Marzani,}
\author[c]{Claudio~Muselli,}
\author[a]{Luca~Rottoli.}

\affiliation[a]{Rudolf Peierls Centre for Theoretical Physics,\\
  1 Keble Road, University of Oxford, OX1 3NP, Oxford, UK}
\affiliation[b]{University at Buffalo, The State University of New York, \\Buffalo, NY 14260-1500, USA}
\affiliation[c]{Dipartimento di Fisica, Universit\`a di Milano and
INFN, Sezione di Milano,\\ Via Celoria 16, I-20133 Milano, Italy}

\emailAdd{marco.bonvini@physics.ox.ac.uk}
\emailAdd{smarzani@buffalo.edu}
\emailAdd{claudio.muselli@mi.infn.it}
\emailAdd{luca.rottoli@physics.ox.ac.uk}

\preprint{
\begin{flushright}
OUTP-16-05P\\
TIF-UNIMI-2016-2
\end{flushright}
}

\abstract{
  We consider the inclusive production of a Higgs boson in gluon-fusion
  and we study the impact of threshold resummation at
  next-to-next-to-next-to-leading logarithmic accuracy (N$^3$LL) on
  the recently computed fixed-order prediction at
  next-to-next-to-next-to-leading order (N$^3$LO).
  We propose a conservative, yet
  robust way of estimating the perturbative uncertainty from missing
  higher (fixed- or logarithmic-) orders.
  We compare our results with two other different methods of estimating the uncertainty from missing higher orders:
  the Cacciari-Houdeau Bayesian approach to theory errors, and the use of algorithms to accelerate
  the convergence of the perturbative series, as suggested by David and Passarino. 
  We confirm that the best convergence happens at $\mur=\muf=\mh/2$, and we conclude that a reliable estimate
  of the uncertainty from missing higher orders on the Higgs cross section at 13~TeV is approximately $\pm4\%$.}

\begin{document}
\maketitle

\section{Introduction}
The major success of the first run of the CERN Large Hadron Collider
(LHC) was the discovery of the Higgs
boson~\cite{Aad:2012tfa,Chatrchyan:2012xdj}.
Run-I studies of this new resonance established its mass and quantum
numbers, as well as its coupling strengths to many Standard Model
particles~\cite{Aad:2015zhl,coupling_comb}. Moreover, measurements of
fiducial cross section and distributions have also been
performed~\cite{Aad:2014tca,Aad:2014lwa,Aad:2015lha,Khachatryan:2015rxa}.
The start of the LHC Run II marked the beginning of the Higgs
precision era.  Now is the time to study the properties of this new
particle in detail so that, nearly fifty years after its proposal, the
Brout-Englert-Higgs mechanism for electroweak symmetry breaking can
undergo its final tests.
One key element in the LHC Run-II rich physics program is the
measurement of the Higgs production cross section and differential
distributions, and their comparison to state-of-the-art theoretical
predictions.
Thus, identifying the different sources of uncertainties and
performing more refined calculations to improve on them is of
paramount importance.
In this paper we consider inclusive production of a Higgs boson in
hadron-hadron collisions, which is chiefly driven by the gluon-fusion
mechanism.
QCD corrections to Higgs production in gluon fusion are known to be
very important, leading to poor convergence properties of the
perturbative series. Indeed, it is known that the next-to-leading
(NLO) contribution corrects the Born cross section by roughly
100\%. Moreover, NNLO corrections are also significant. Recently, a
milestone calculation of third-order perturbative coefficient was
performed~\cite{Anastasiou:2014vaa,Anastasiou:2014lda,Anastasiou:2015ema,Anastasiou:2016cez}.
N$^3$LO corrections appear to be small, indicating that the
perturbative series is finally manifesting convergence.

Threshold resummation is a reorganization of the perturbative
expansion that accounts for logarithmically enhanced contributions to
all-orders in the strong coupling $\as$. In the same way as
fixed-order calculations can be performed at different (N$^k$LO)
accuracy, resummed calculations can be systematically improved by
including next$^k$-to-leading logarithmic (N$^k$LL) corrections.
Threshold resummation for Higgs production is currently known to
N$^3$LL~\cite{Bonvini:2014joa,Catani:2014uta,Bonvini:2014tea,Schmidt:2015cea}.
In this work we update the results of Refs.~\cite{Bonvini:2014joa},
in view of the recently computed N$^3$LO
result~\cite{Anastasiou:2014vaa,Anastasiou:2014lda,Anastasiou:2015ema,Anastasiou:2016cez}.
We discuss the impact of the resummation on the central value and we
suggest a robust way of estimating the theoretical uncertainty due to
missing higher orders, which goes beyond traditional scale variation.

The paper is organized as follows. We start in Sect.~\ref{sec:n3lo} by
reviewing the results of
Refs.~\cite{Anastasiou:2014vaa,Anastasiou:2014lda,Anastasiou:2015ema,Anastasiou:2016cez}
and using them to assess the validity of the fixed-order approximation
based on analyticity properties of perturbative coefficient
functions of Refs.~\cite{Ball:2013bra,Bonvini:2014jma}.  In Sect.~\ref{sec:resummation}, we then
review basic results in threshold resummation as well as describe the
improvements suggested in
Ref.~\cite{Bonvini:2014joa}. Sect.~\ref{sec:matching} contains the main
results of this study, namely the matched N$^3$LO+N$^3$LL cross
section and a detailed study of the perturbative uncertainty. Finally,
in Sect.~\ref{sec:missing_HO} we compare our findings to different
methods to assess the size of missing higher orders, namely the
Cacciari-Houdeau
method~\cite{Cacciari:2011ze,Forte:2013mda,Bagnaschi:2014wea} and the
application of convergence acceleration algorithms to the perturbative
series, as suggested by David and Passarino~\cite{David:2013gaa}.

\section{The N$^3$LO cross section}\label{sec:n3lo}

In order to fix the notation, we write the hadron-level cross section for the production
of a Higgs boson with mass $\mh$ in hadron-hadron collisions as
\beq\label{eq:xs}
\sigma(\tau,\mh^2,\mt^2) = \tau
\sum_{ij}\int_\tau^1 \frac{dz}{z}\,\Lum_{ij}\(\frac{\tau}{z},\muf^2\)
\frac1z \hat\sigma_{ij}\(z, \mh^2, \mt^2, \as(\mur^2),\frac{\mh^2}{\muf^2},\frac{\mh^2}{\mur^2}\),
\qquad
\tau=\frac{\mh^2}{s},
\eeq
where $\Lum_{ij}(z,\mu^2)$ is a parton luminosity
\beq\label{eq:lum}
\Lum_{ij}(z,\mu^2) = \int_z^1 \frac{dx}x\, f_i\(\frac zx,\mu^2\) f_j(x,\mu^2),
\eeq
and $i,j$ run over all parton flavours.
In the dominant gluon-fusion production mechanism, the Higgs is generated by the fusion of two gluons through a quark loop.
For simplicity, we have only considered a top quark with mass $\mt$ running in the loop.
Moreover, for ease of notation, the dependence on factorization scale $\muf$ and renormalization scale $\mur$ is often left understood.

The partonic cross section $\hat\sigma_{ij}$ is then related to the dimensionless coefficient function $C_{ij}$ by
\beq\label{eq:partonic_xs}
\hat\sigma_{ij}(z,\mh^2,\mt^2,\as) = z\,\sigma_0(\mh^2,\mt^2)\,C_{ij}(z, \mh^2,\mt^2,\as).
\eeq
where $\sigma_0$ is such that the coefficient function has the perturbative expansion in the strong coupling $\as$
\begin{align}\label{eq:Cpert}
C_{ij}(z,\mh^2,\mt^2,\as) &= \delta_{ig}\delta_{jg}\delta(1-z) + \as\,C_{ij}^{(1)}(z,\mh^2,\mt^2) + \as^2\,C_{ij}^{(2)}(z,\mh^2,\mt^2)
\nonumber \\&+ \as^3\,C_{ij}^{(3)}(z,\mh^2,\mt^2) + \Ord(\as^4).
\end{align}
While the NLO coefficient $C_{ij}^{(1)}$ is known
exactly~\cite{Spira:1995rr} and $C_{ij}^{(2)}$ is known as an
expansion in $\mh^2/\mt^2$ matched to the (full-theory) small-$z$
limit~\cite{Marzani:2008az,Harlander:2009bw,Harlander:2009mq,Harlander:2009my,Pak:2009bx,Pak:2009dg},
the third order coefficient has been computed only recently in the
large-$\mt$ effective
theory~\cite{Anastasiou:2014vaa,Anastasiou:2014lda,Anastasiou:2015ema,Anastasiou:2016cez}.

In the large-$\mt$ effective field theory (EFT), the heavy top is integrated out and consequently
the dependence on the top mass only appears in a Wilson coefficient squared $W$.
The EFT is usually improved with a rescaling of the cross section by the ratio of the exact LO over the LO in the EFT,
leading to the so-called rescaled effective theory (rEFT), which is known to be a very good approximation for $\mh\lesssim\mt$
and for not too high collider energies, essentially because the large-$s$, i.e.\ small-$z$,
behaviour of the EFT coefficient functions is double-logarithmic~\cite{Hautmann:2002tu},
while the full theory only exhibits single high-energy logarithms~\cite{Marzani:2008az}.
In the EFT, the coefficient function further factorizes
\beq\label{eq:EFT}
C_{ij}(z,\mh^2,\mt^2,\as) = W(\mh^2,\mt^2) \,\tilde C_{ij}(z,\as),
\eeq
where $\tilde C_{ij}$ has an expansion analogous to $C_{ij}$, Eq.~\eqref{eq:Cpert}, and $W=1+\Ord\(\as\)$.
Thus, in the EFT, the coefficient $C^{(3)}_{ij}$ is given by
\beq\label{eq:C3eft}
C^{(3)}_{ij} = \tilde C^{(3)}_{ij} + W^{(1)} \tilde C^{(2)}_{ij}  + W^{(2)} \tilde C^{(1)}_{ij}  + W^{(3)} \delta_{ig}\delta_{jg},
\eeq
where $W^{(k)}$ are the coefficients of the expansion of the Wilson coefficient squared $W$ in powers of $\as$.
The lower order coefficients $\tilde C^{(1)}_{ij}$ and $\tilde C^{(2)}_{ij}$ are fully known,
while $\tilde C^{(3)}_{ij}$ is the recently computed N$^3$LO contribution published in Ref.~\cite{Anastasiou:2016cez}.

For its computation, the third order coefficient $\tilde C^{(3)}_{ij}$ has been decomposed into contributions
proportional to $\ln^k(1-z)$ with $k=0,\dots, 5$. For $k=3, 4$ and 5 the exact result is known~\cite{Anastasiou:2014lda},
while for lower powers 0, 1 and 2, the result has been expressed in terms of  a soft expansion in $1-z$ up to order $(1-z)^{37}.$\footnote
{The difference between the exact expression and the soft expansion for the three highest powers of $\ln(1-z)$ is totally negligible.}
While not included in the result of Ref.~\cite{Anastasiou:2014lda}, we stress that the leading small-$z$ logarithm,
$\frac1z\ln^5z$, is known from small-$z$ resummation~\cite{Hautmann:2002tu} (see App.~\ref{app:ggHiggs} for details).

The construction of $C^{(3)}_{ij}$, Eq.~\eqref{eq:C3eft}, has then some degree of arbitrariness.
The formal accuracy is set by the soft expansion of $\tilde C^{(3)}_{ij}$, so in principle one could
take a soft expansion of all its ingredients, in particular the lower orders $\tilde C^{(1)}_{ij}$ and $\tilde C^{(2)}_{ij}$.
Alternatively, one can retain as much available information as possible, thereby using the exact expressions
for $\tilde C^{(1)}_{ij}$ and $\tilde C^{(2)}_{ij}$ and adding the leading small-$z$ logarithm.
These (and any other intermediate) options are all perfectly valid and the difference among them could be considered a measure of the uncertainty related to the soft expansion.
We have found that the residual uncertainty on the soft expansion of $C^{(3)}_{ij}$ mostly comes from the small-$z$
logarithms. 
Thus, for all phenomenological applications where the small-$z$
logarithms are negligible, the soft expanded result for $C^{(3)}_{ij}$
(in any practical incarnation) is perfectly reliable and accurate, in
full agreement with the analysis of Ref.~\cite{Anastasiou:2014lda}.
On the other hand, the uncertainty related to the soft expansion
becomes larger as we increase the centre-of-mass energy, leading to
sizeable effects at energy scales of Future Circular Colliders
(FCC), $\sqrt{s}\simeq100$~TeV. Furthermore, even if we controlled the full $z$ dependence of
$C^{(3)}_{ij}$, the EFT result itself becomes unreliable at very large
energies because of the appearance of double logarithmic
contributions, as previously discussed.

Therefore, rather than improving the EFT result with EFT small-$z$
logarithms, one can try to improve it with finite $\mt$ effects.  One
way to do it is including those $\mt$-dependent contributions which
are predicted by all-order calculations in the threshold and
high-energy limits. As we will see explicitly in
Sect.~\ref{sec:resummation}, the $\mt$ dependence of the soft
logarithms appears in a factorized form, and it is therefore possible
to account for the exact $\mt$ dependence of all the logarithmic
terms, i.e.\ all the plus-distributions, that appear at N$^3$LO.
The situation in the opposite, small-$z$, limit is less satisfying
because the resummation is known strictly speaking only at the first
non-trivial order~\cite{Marzani:2008az,Ball:2013bra}, although a large
class of subleading running-coupling corrections can also be
resummed~\cite{Altarelli:2005ni,Altarelli:2008aj}.
Both effects have been implemented\footnote
{We thank Achilleas Lazopoulos for providing the coefficients of the soft expansion of the N$^3$LO result.}
in the code \texttt{ggHiggs}, version 3.1, publicly available from the webpage~\cite{ggHiggs}.
Further details on the \texttt{ggHiggs} implementation are given in App.~\ref{app:ggHiggs}.

\begin{figure}[t]
  \centering
  \includegraphics[width=0.495\textwidth,page=1]{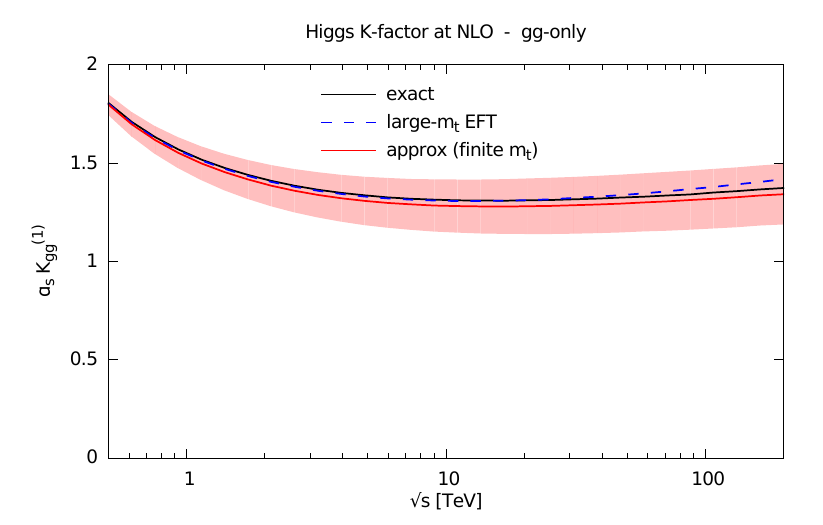}
  \includegraphics[width=0.495\textwidth,page=2]{images/paper_hadr_Kfact_new_PDF4LHC15.pdf}\\
  \includegraphics[width=0.7\textwidth,page=4]{images/paper_hadr_Kfact_new_PDF4LHC15.pdf}
  \caption{Perturbative $K$-factors at NLO (top left), NNLO (top right) and N$^3$LO (bottom) as a function of the collider energy,
    for $\mh=125$~GeV.}
  \label{fig:Kfact}
\end{figure}
In order to illustrate our findings, we now focus on the dominant $gg$ channel,
and consider the ``$n$-th order $K$-factor'' $K_{gg}^{(n)}$,
defined as the contribution to the $K$-factor coming from $C_{gg}^{(n)}$ only, namely
\beq\label{eq:K3}
K_{gg}^{(n)} = \Lum_{gg}^{-1}(\tau)
\int_\tau^1 \frac{dz}{z}\,\Lum_{gg}\(\frac{\tau}{z}\)\,
C_{gg}^{(n)}(z, \mh^2, \mt^2)
\eeq
such that
\beq
K_{gg} = 1 + \as K_{gg}^{(1)} + \as^2 K_{gg}^{(2)} + \as^3 K_{gg}^{(3)} + \ldots
\eeq
In Fig.~\ref{fig:Kfact} we show the results for the NLO, NNLO and N$^3$LO $K$-factors
as a function of the collider energy $\sqrt{s}$,
for fixed Higgs mass $\mh=125$~GeV, and top pole mass $\mt=172.5$~GeV.
We use the \texttt{PDF4LHC15\_nnlo\_100} PDF set~\cite
{Butterworth:2015oua,Carrazza:2015aoa,Ball:2014uwa,Harland-Lang:2014zoa,Dulat:2015mca}.
Results obtained in the large-$\mt$ EFT are shown in dashed blue,
while solid black curves also contain finite-$\mt$ corrections (we
remind the Reader that the $\mt$ dependence is fully accounted for at
NLO, while it is treated as a power expansion at NNLO). In all cases,
differences between large-$\mt$ and finite-$\mt$ is small, but it
increases with $\sqrt{s}$, as expected.

At N$^3$LO, we computed the EFT result using the exact expressions for $\tilde C^{(1)}_{ij}$ and $\tilde C^{(2)}_{ij}$
in Eq.~\eqref{eq:C3eft}, while $\tilde C^{(3)}_{ij}$ is soft-expanded.\footnote
{The effect of soft-expanding the lower order is rather mild, while the effect of including the leading
small-$z$ logarithm is comparable in size to the difference between dashed blue and solid black curves in Fig.~\ref{fig:Kfact}.}
The ``$\mt$-improved EFT'' curve (black solid) is constructed as described above:
in particular, the coefficients of all the large-$z$ logarithms have the correct $\mt$-dependence, while in the opposite, small-$z$, limit we have included the appropriate leading logarithmic term $\frac1z\ln^2z$.

In all plots of Fig.~\ref{fig:Kfact} we additionally show the approximate result of
Ref.~\cite{Ball:2013bra,Bonvini:2014jma} in red, with its uncertainty, in order to assess the validity of that approach.
The approximation is based on a combination of the soft and high-energy behaviours, including finite $\mt$ effects.
The soft part of the approximation, which gives the dominant contribution,
presents several improvements with respect to standard soft approximations (see e.g.~\cite{Moch:2005ky});
in particular, two versions of the soft approximation, denoted soft$_1$ and soft$_2$, have been considered,
and the area between the two has been considered as the uncertainty on the soft result, using the average as central prediction.
This improved soft definition has been later extended to all orders in Ref.~\cite{Bonvini:2014joa};
in that work, two of us noticed that soft$_2$ (denoted A-soft$_2$ there) performed much better than (A-)soft$_1$.
Since then, we decided to centre the approximation on soft$_2$, symmetrizing the difference between soft$_1$ and soft$_2$
about soft$_2$ to get the uncertainty (which is then twice as large as in the original version).
This is how we now compute the (soft part of the) error in Fig.~\ref{fig:Kfact}. This prescription was also used for the prediction published in Ref.~\cite{Aad:2015lha}.

As known from Ref.~\cite{Ball:2013bra}, the approximation is very accurate at NLO and NNLO, the exact result lying well
within the uncertainty band, as shown in the first two panels of Fig.~\ref{fig:Kfact}.
In the third panel, we show for the first time the approximate prediction of the N$^3$LO compared with the full EFT result of Ref.~\cite{Anastasiou:2016cez} (in dashed blue) and to the $\mt$-improved EFT (in solid black).

As expected, at lower collider energies, where the process is closer to threshold, the approximation is very good and perfectly
compatible with the exact result within uncertainties. Note that at these energy scales the missing terms
beyond the soft expansion at N$^3$LO are completely negligible, so the full N$^3$LO result can be regarded as exact.
At very high (FCC) energies, the comparison between EFT (blue) and approximation (red) is not significant.
Perhaps surprisingly, we also find that our approximation and the $\mt$-improved EFT prediction differ
at very high energy (although they are still compatible within errors) despite having the same leading logarithmic behaviour. 
The difference is due to subleading high-energy terms $\frac1z\ln z$
and $\frac1z$, which are included in the approximate result because
they are driven by the resummation of the splitting functions~\cite{Ball:2013bra},
but not in the $\mt$-improved EFT because they are not fully determined. The effect of these subleading terms is rather large
(larger than at NNLO) and a full determination of at least the $\frac1z\ln z$
contribution is highly desirable.

Finally, we see that in the intermediate region, relevant for LHC, our approximation works less well and the full result
lies at the lower edge of the uncertainty band of the approximate result.
This seems to suggest that, at N$^3$LO, the contributions which are neither soft nor high-energy are more important than at previous orders,
a fact which was not taken into account when estimating the uncertainty from these terms in Ref.~\cite{Ball:2013bra}.
Nevertheless, in the $gg$ channel considered here, the overall agreement of the approximate result with the full result remains rather good.

\section{Threshold resummation(s)} \label{sec:resummation}

Soft-gluon resummation is usually performed in Mellin ($N$) space,
where the multiple gluon emission phase-space factorizes.
The cross section Eq.~(\ref{eq:xs}) has a simpler structure in Mellin space:
\beq\label{eq:xs_mell}
\int_0^1d\tau\,\tau^{N-1}\frac{\sigma(\tau,\mh^2)}\tau =\sigma_0(\mh^2,\mt^2)\, \Lum(N)\, C\(N,\as\),
\eeq
where we have used the same symbols, with different arguments, for a function and its Mellin transform.
Note that threshold resummation only affects the $gg$ channel: we therefore suppress the flavour indices and implicitly focus on the $gg$ channel.
We will later comment on the role of the quark channels. 
The $N$-space resummed coefficient function
has the form (see~\cite{Bonvini:2014joa} and reference therein):
\begin{align}
\label{eq:Cres}
C_{\rm res}\(N,\as\)&=\bar g_0\(\as,\muf^2\) \exp\bar\Sud(\as,N),\\
\bar\Sud(\as,N)
&= \int_0^1dz\, \frac{z^{N-1}-1}{1-z}
\( \int_{\muf^2}^{\mh^2\frac{(1-z)^2}{z}} \frac{d\mu^2}{\mu^2} 2A\(\as(\mu^2)\) + D\(\as([1-z]^2\mh^2)\) \),\label{eq:Sbar}\\
\bar g_0(\as,\muf^2) &= 1+\sum_{k=1}^\infty \bar g_{0,k}(\muf^2) \as^k,\\
A(\as)&=\sum_{k=1}^\infty A_k\as^k, \qquad
D(\as)=\sum_{k=1}^\infty D_k\as^k,\label{eq:A,D}
\end{align}
where $\bar g_0(\as,\muf^2)$ does not depend on $N$. We note that in the full theory, all the top-mass dependence is in $\bar g_0$. 
Furthermore, under the rEFT assumption, its expression factorizes as
\beq
\bar g_0(\as,\muf^2) = W(\mh^2,\mt^2,\muf^2) \, \tilde{\bar{g}}_0(\as,\muf^2)
\eeq
where now $\tilde{\bar g}_0(\as,\muf^2)$ does not depend on the top mass.
Note that we have restored explicit scale dependence and we have chosen
the factorization scale $\muf$ as the scale of the running coupling $\as=\as(\muf^2)$.
The three-loop coefficients of $A(\as)$ and $D(\as)$ have been known
for a while (see for instance Refs.~\cite{Moch:2005ba, Moch:2005ky,Laenen:2005uz}), while the $\Ord
\(\as^3 \)$ contribution to $\tilde{\bar{g}}_0$ has been recently computed~\cite{Anastasiou:2014vaa}.
The four-loop contribution to
$A(\as)$, which is needed to achieve full
N$^3$LL accuracy, is unknown. However, a Pad\'e
estimate~\cite{Moch:2005ba} can suggest the size of its value, and a
numerical analysis shows that its impact in a resummed result is
essentially negligible.

The integrals in Eq.~\eqref{eq:Sbar} can be computed at any finite logarithmic accuracy by using
the explicit solution of the running coupling,
in terms of $\as$ at a given reference scale, which we can also choose to be $\muf$ in first place.
At this point we have a result which depends on a single scale $\muf$, with $\as$ always computed at $\muf$
(note that, while the $\muf$ dependence of $\bar\Sud$ is explicit,
the one of $\bar g_0$ can be recovered by imposing $\muf$-independence of the full cross section).
In order to write the result in a canonical way, we further evolve $\as$ from $\muf$ to $\mur$ using the explicit solution of the running coupling equation
at sufficiently high order, and propagating the resulting logarithms in the various terms at each fixed-order (in $\bar g_0$)
and logarithmic-order (in $\bar\Sud$) accuracy.
Then, the final result explicitly depends on both $\mur$ and $\muf$.

The computation of the integrals in Eq.~\eqref{eq:Sbar} is rather cumbersome when performed exactly.
The resulting expression was called A-soft in Ref.~\cite{Bonvini:2014joa}.
The computation is much simpler when performed in the large-$N$ limit, where the result of the integrals
is written as a function of $\ln N$ only. We call the result in this limit $N$-soft.
Explicit expressions for $\bar\Sud$ in the $N$-soft limit up to N$^3$LL are given in Ref.~\cite{Moch:2005ba}\footnote
{To be precise, the expressions in Ref.~\cite{Moch:2005ba} are for the logarithmic part of the exponent, and not for the $N$-independent terms.}
with full $\muf$ and $\mur$ dependence.

In Ref.~\cite{Bonvini:2014joa} two of us proposed a variant of the $N$-soft resummation based on the simple replacement
\beq
\ln N \to \psi_0(N),
\eeq
$\psi_0(N)$ being the Euler digamma function. This prescription was called $\psi$-soft.
It has the advantage that it reproduces
the function $\bar\Sud$ up to corrections of $\Ord(1/N^2)$, while $N$-soft reproduces
$\bar\Sud$ only up to $\Ord(1/N)$ corrections.
Note that this does not mean that $\psi$-soft resummation is accurate at next-to-soft (NS) $\Ord(1/N)$ level,
because the original expression Eq.~\eqref{eq:Cres} was not.
However, as pointed out in Ref.~\cite{Bonvini:2014joa}, a class of NS terms can be predicted to all orders by
adding a ``collinear improvement'' to Eq.~\eqref{eq:Sbar}. This is achieved by recalling that the function $A(\as(\mu^2))/(1-z)$
is nothing but the divergent part (in the $z\to1$ limit) of the Altarelli-Parisi splitting function $P_{gg}(z,\as)$.
One can therefore keep more terms in the soft-expansion of $P_{gg}$ about $z=1$.
As shown in Ref.~\cite{Bonvini:2014joa}, by using the LO gluon splitting function up to order $(1-z)^{k-1}$
one accounts for the leading-logarithmic (LL) N$^k$S terms correctly to all orders.

In Ref.~\cite{Bonvini:2014joa} two variants of the collinear improvement,  AP1 and AP2, were considered,
obtained by expanding the LO $P_{gg}$ to first and second order in $1-z$, respectively. 
Since these corrections correspond to extra powers of $z$, the effect is to shift the
value of $N$. Extending the shifts to the whole $\bar\Sud$ function, we therefore have\footnote
{A study of different resummation prescriptions was also performed in Ref.~\cite{Anastasiou:2016cez}.
However, the definition of AP2 (and hence of $\psi$-soft AP2) used in that study differs from the one employed here,
which is based on Ref.~\cite{Bonvini:2014joa}.}
\begin{align}
  \label{eq:AP}
  &\text{AP1:} &\bar\Sud(\as,N)&\to \bar\Sud(\as,N+1) \\
  &\text{AP2:} &\bar\Sud(\as,N)&\to 2\bar\Sud(\as,N) - 3\bar\Sud(\as,N+1) + 2\bar\Sud(\as,N+2).
\end{align}
We have verified that AP2 leads to more reliable results, in the sense that expanding the $\psi$-soft resummed expression
with AP2 to fixed-order reproduces to a good accuracy the exact results
(see Ref.~\cite{Ball:2013bra,Muselli:2015kba}, and discussion in Sect.~\ref{sec:n3lo} which extends the validation to the third order).
In fact, comparing AP2 with AP1 also allows to estimate the uncertainty due to missing $1/N$ terms, and constructing an uncertainty
from the difference between AP2 and AP1 was indeed successful
(see again Ref.~\cite{Ball:2013bra} and Sect.~\ref{sec:n3lo}).

We finally turn to discussing another source of uncertainty at resummed level coming from subleading logarithmic terms.
The function $\bar\Sud$, Eq.~\eqref{eq:Sbar}, contains on top of logarithmically enhanced contributions
(terms which grow logarithmically at large $N$)
also constant ($N$-independent) terms, analogous to those included in $\bar g_0$.
In standard $N$-soft resummation (see e.g.\ Refs.~\cite{Catani:2003zt,deFlorian:2012yg, Catani:2014uta})
all the constants are usually removed from the exponent and collected
in the function in front,
\beq\label{eq:constantsg0}
C_{\rm res}\(N,\as\)= g_0\(\as,\muf^2\) \exp\Sud(\as,N),
\eeq
which then changes name to $g_0$ (without bar; consequently also $\bar\Sud$ changes into a new function $\Sud$ which contains only logarithmic terms).
Alternatively, all constants can be moved into the exponent~\cite{Bonvini:2014joa},
\beq\label{eq:constantsExp}
C_{\rm res}\(N,\as\)= \exp\[\ln\bar g_0\(\as,\muf^2\) + \bar\Sud(\as,N) \],
\eeq
where $\ln\bar g_0$ is meant to be expanded to the appropriate order (which is $\Ord(\as^3)$ for N$^3$LL accuracy).\footnote
{We recall that sometimes the N$^3$LL accuracy we refer to here is (perhaps more correctly) referred to as N$^3$LL$^\prime$.}
Up to the working logarithmic accuracy, the position of the constants does not make any difference.
However, beyond the working logarithmic accuracy, moving constants produces, by interference, different subleading term.
Therefore, one can consider Eq.~\eqref{eq:constantsg0} and Eq.~\eqref{eq:constantsExp} as two opposite options
which treat in a maximally different way these subleading terms, and use them to assign an uncertainty
to the default (most natural) expression Eq.~\eqref{eq:Cres}.
Note that, since constants are known to play an important role for Higgs production~\cite{Ahrens:2008qu,Berger:2010xi,Stewart:2013faa},
these variations provide a robust way to estimate the perturbative uncertainty.

\section{Threshold resummation at N$^3$LL and its uncertainties} \label{sec:matching}
\label{sec:results}

Having described the various prescriptions available for the threshold resummation, we now 
move to a description of how we propose to use them to improve the central value and most importantly
the uncertainty from missing higher (fixed- or logarithmic-) orders for the inclusive Higgs cross section.

First, we discuss how threshold resummation is matched to a fixed-order calculation.
The coefficient function $C_{\rm res}(N,\as)$ contains all orders in $\as$ but it is accurate only in the soft limit.
Assuming we have available the exact result for the coefficient function up to $\Ord(\as^k)$,
to maximally use information from both exact fixed order and soft all orders, one should use the fixed-order result
up to (and including) $\Ord(\as^k)$, and the resummed result for the remaining terms from $\as^{k+1}$ onwards.
Matching is achieved by simply adding together the two calculations and subtracting the expansion of the resummation to $\Ord(\as^k)$, in order to avoid double counting. 
We then define
\beq\label{eq:DeltaC}
\Delta_k C_{\rm res}(N,\as) = C_{\rm res}(N,\as) - \sum_{i=0}^k \as^i\, C_{\rm res}^{(i)}(N)
\eeq
being $C_{\rm res}^{(j)}(N)$ the coefficients of the expansion of $C_{\rm res}(N,\as)$ in powers of $\as$.
Therefore, the matched cross section is written as
\beq
\sigma_{\text{N$^k$LO+N$^j$LL}} = \sigma_{\text{N$^k$LO}} + \Delta_k\sigma_{\text{N$^j$LL}},
\eeq
where
\beq\label{eq:Deltasigma}
\Delta_k\sigma_{\text{N$^j$LL}} = \tau\, \sigma_0(\mh^2,\mt^2) \int_{c-i\infty}^{c+i\infty} \frac{dN}{2\pi i}\,\tau^{-N}\, \Lum_{gg}(N)\, \Delta_k C_{\text{res,N$^j$LL}}(N,\as)
\eeq
is the inverse Mellin transform of the gluon-gluon luminosity times $\Delta_k C_{\rm res}$, Eq.~\eqref{eq:DeltaC}.
By construction, $\Delta_k\sigma_{\text{N$^j$LL}}$ Eq.~\eqref{eq:Deltasigma} only contains higher-order corrections to $\sigma_{\text{N$^k$LO}}$.

The computation of $\Delta_k\sigma_{\text{N$^j$LL}}$ Eq.~\eqref{eq:Deltasigma} is done through the public code \texttt{TROLL}~\cite{TROLL},
formerly \texttt{ResHiggs}, which changed name after the inclusion of Drell-Yan and DIS resummation~\cite{Bonvini:2015ira}.
Some details on the validation of the code are given in App.~\ref{app:TROLL}.
As the name implies (the meaning of \texttt{TROLL} is \texttt{TROLL Resums Only Large-x Logarithms}), the code does not compute the fixed
order, but only the resummed contribution, Eq.~\eqref{eq:Deltasigma}, so the fixed order has to be supplied by an external code.
In this work, we use the code \texttt{ggHiggs}~\cite{ggHiggs}, where we
have implemented the new N$^3$LO result~\cite{Anastasiou:2014vaa,Anastasiou:2014lda,Anastasiou:2015ema,Anastasiou:2016cez}.
We stress that any other fixed-order code can be used (e.g., future versions of \texttt{ihixs}~\cite{ihixs}),
provided the same setting is used in both codes.
A new version of \texttt{TROLL}, v3.1, which interfaces directly with \texttt{ggHiggs} v3.1
is publicly available from the webpage~\cite{TROLL}.

For simplicity, and for disentangling effects coming from different sources, we work in the clean environment of the
(rescaled) large-$\mt$ effective theory (rEFT), using the top mass $\mt=172.5$~GeV in the pole scheme.
We take the Higgs mass to be $\mh=125$~GeV, and use the \texttt{PDF4LHC15\_nnlo\_100} PDF set~\cite
{Butterworth:2015oua,Carrazza:2015aoa,Ball:2014uwa,Harland-Lang:2014zoa,Dulat:2015mca}.
As far as $\as$ is concerned, we take $\as(m_Z^2)=0.118$ from the PDF set and evolve it at four loops to $\mur$.
We focus on LHC at $\sqrt{s}=13$~TeV first, and consider different collider energies later.

To show the stability of the resummed result, we consider four options for the central common factorization and renormalization scale $\mu_0$:
\beq\label{eq:centralscales}
\mu_0 = \mh/4,\qquad
\mu_0 = \mh/2,\qquad
\mu_0 = \mh,\qquad
\mu_0 = 2\mh.
\eeq
We then vary the scales $\mur$ and $\muf$ about $\mu_0$ by a factor of 2 up and down,
keeping the ratio $\mur/\muf$ never larger than 2 or smaller than $1/2$: we call this a canonical 7-point scale variation.
This results in a total of 16 different combination of scales.
Results at fixed order for these scales are presented in Tab.~\ref{tab:FO}.

\begin{table}[t]
  \centering
  \begin{tabular}{cc|cccc}
    $\muf/\mh$ & $\mur/\mh$ & LO & NLO & NNLO & N$^3$LO \\
    \midrule
$   4$ & $   4$ & $10.2$ & $24.1$ & $35.1$ & $41.8$ \\
$   4$ & $   2$ & $12.2$ & $28.0$ & $39.1$ & $44.7$ \\
$   2$ & $   4$ & $10.0$ & $23.6$ & $34.6$ & $41.4$ \\
$   2$ & $   2$ & $11.9$ & $27.5$ & $38.6$ & $44.3$ \\
$   2$ & $   1$ & $14.4$ & $32.3$ & $43.0$ & $46.8$ \\
$   1$ & $   2$ & $11.4$ & $26.8$ & $38.0$ & $43.9$ \\
$   1$ & $   1$ & $13.8$ & $31.6$ & $42.4$ & $46.5$ \\
$   1$ & $ 1/2$ & $17.0$ & $37.9$ & $47.0$ & $48.2$ \\
$ 1/2$ & $   1$ & $13.0$ & $30.7$ & $41.8$ & $46.2$ \\
$ 1/2$ & $ 1/2$ & $16.0$ & $36.9$ & $46.5$ & $48.1$ \\
$ 1/2$ & $ 1/4$ & $20.3$ & $45.3$ & $50.7$ & $48.0$ \\
$ 1/4$ & $ 1/2$ & $14.7$ & $35.7$ & $46.1$ & $48.0$ \\
$ 1/4$ & $ 1/4$ & $18.6$ & $44.2$ & $50.7$ & $48.1$ \\
$ 1/4$ & $ 1/8$ & $24.4$ & $56.3$ & $52.6$ & $44.5$ \\
$ 1/8$ & $ 1/4$ & $16.2$ & $42.6$ & $51.0$ & $46.4$ \\
$ 1/8$ & $ 1/8$ & $21.3$ & $55.2$ & $54.1$ & $40.6$ \\
  \end{tabular}
  \caption{Fixed-order cross sections (in pb) as a function of the scales $\muf$ and $\mur$ over a wide range,
    for $\mh=125$~GeV at LHC with $\sqrt{s}=13$~TeV in the rEFT.
    The numerical integration error on all results is below the number of digits shown.
    The values of $\as$ at the six renormalization scales $\mur=\{4\mh,2\mh,\mh,\mh/2,\mh/4,\mh/8\}$
    are $\as(\mur)=\{0.0940,0.1024,0.1126,0.1252,0.1409,0.1614\}$.}
  \label{tab:FO}
\end{table}

For each pair of scales, we then compute the resummed contribution with \texttt{TROLL} for different variants of the resummation.
We consider
\begin{itemize}
\item standard $N$-soft, with all constants in $g_0$, Eq.~\eqref{eq:constantsg0}, without collinear improvement;
\item $\psi$-soft with AP2, 
\begin{itemize}
\item  with default choice for the constants, Eq.~\eqref{eq:Cres};
\item  with all constants in $g_0$, Eq.~\eqref{eq:constantsg0};
\item  with all constants in the exponent, Eq.~\eqref{eq:constantsExp};
\end{itemize}
\item $\psi$-soft with AP1, 
\begin{itemize}
\item  with default choice for the constants, Eq.~\eqref{eq:Cres};
\item  with all constants in $g_0$, Eq.~\eqref{eq:constantsg0};
\item  with all constants in the exponent, Eq.~\eqref{eq:constantsExp};
\end{itemize}
\end{itemize}
The resummed results at N$^3$LO+N$^3$LL, obtained as the sum of the last column of Tab.~\ref{tab:FO}
and the resummation contribution $\Delta_3\sigma_{\text{N$^3$LL}}$ computed with \texttt{TROLL},
are shown in Tab.~\ref{tab:res}.

\begin{table}[t]
  \centering
\begin{tikzpicture}
\node (table) {%
  \begin{tabular}{cc|rrrrrrr}
    & & \multicolumn{6}{c|}{$\psi$-soft} & \\
    & & \multicolumn{2}{c|}{default} & \multicolumn{2}{c|}{constants in exp} & \multicolumn{2}{c|}{constants in $g_0$} & \\
    $\muf/\mh$ & $\mur/\mh$ & AP2 & AP1 & AP2 & AP1 & AP2 & AP1 & \multicolumn{1}{|c}{$N$-soft}\\
    \midrule
$   4$ & $   4$ & $56.8$ & $66.0$ & $56.8$ & $66.0$ & $51.2$ & $58.7$ & $49.4$ \\
$   4$ & $   2$ & $55.1$ & $62.3$ & $54.9$ & $62.0$ & $52.2$ & $58.6$ & $50.5$ \\
$   2$ & $   4$ & $53.2$ & $57.2$ & $53.7$ & $57.9$ & $48.2$ & $51.4$ & $46.0$ \\
$   2$ & $   2$ & $52.9$ & $56.0$ & $52.7$ & $55.8$ & $49.9$ & $52.5$ & $47.9$ \\
$   2$ & $   1$ & $51.2$ & $53.0$ & $50.9$ & $52.6$ & $50.5$ & $52.1$ & $48.9$ \\
$   1$ & $   2$ & $50.2$ & $50.4$ & $50.6$ & $50.9$ & $47.6$ & $47.7$ & $45.6$ \\
$   1$ & $   1$ & $50.1$ & $50.1$ & $49.8$ & $49.8$ & $49.1$ & $49.0$ & $47.5$ \\
$   1$ & $ 1/2$ & $48.5$ & $48.3$ & $48.3$ & $48.0$ & $49.1$ & $48.8$ & $48.3$ \\
$ 1/2$ & $   1$ & $48.4$ & $47.4$ & $48.8$ & $47.7$ & $47.6$ & $46.6$ & $46.3$ \\
$ 1/2$ & $ 1/2$ & {\color{red}$48.5$} & $48.0$ & $48.3$ & $47.8$ & $48.6$ & $48.1$ & $48.0$ \\
$ 1/2$ & $ 1/4$ & $47.0$ & $47.1$ & $47.1$ & $47.2$ & $47.7$ & $47.7$ & $47.9$ \\
$ 1/4$ & $ 1/2$ & $47.8$ & $47.4$ & $48.2$ & $47.7$ & $48.0$ & $47.6$ & $47.6$ \\
$ 1/4$ & $ 1/4$ & $47.7$ & $48.0$ & $47.6$ & $47.9$ & $48.0$ & $48.2$ & $48.2$ \\
$ 1/4$ & $ 1/8$ & $44.7$ & $45.1$ & $45.4$ & $45.7$ & $44.6$ & $45.0$ & $44.9$ \\
$ 1/8$ & $ 1/4$ & $45.5$ & $46.1$ & $46.1$ & $46.6$ & $46.2$ & $46.6$ & $46.5$ \\
$ 1/8$ & $ 1/8$ & $41.0$ & $40.9$ & $41.4$ & $41.2$ & $40.9$ & $40.8$ & $40.9$ \\
  \end{tabular}};
\draw [red, dashed,rounded corners]
  ($(table.south west) !.225! (table.north west)  !.26! (table.south east)$)
  rectangle 
  ($(table.south east)  !.595! (table.north east)  !.13! (table.south west)$);
\end{tikzpicture}
  \caption{Resummed cross sections (in pb) at N$^3$LO+N$^3$LL cross section for the different prescriptions.
    Scales and settings as in Tab.~\ref{tab:FO}.}
  \label{tab:res}
\end{table}

It is well known~\cite{Catani:2003zt} that
threshold resummation introduces a dependence on the factorization
scale which can be larger than what is obtained in fixed-order
perturbation theory.
This is essentially due to the fact that threshold resummation predicts only the (dominant) $gg$ channel,
while factorization scale dependence is compensated among different channels,
as DGLAP evolution mixes quarks and gluons.
Moreover, at fixed order the factorization scale dependence for Higgs production is very mild
(and much milder than renormalization scale dependence, see Tab.~\ref{tab:FO}),
so the factorization scale dependence of the resummed result is visibly larger.
The quark channels, of which $qg$ gives the most important contribution, give rise to
logarithmic terms that are suppressed by $1/N$, in the large $N$ limit.
Including a prediction of this channel to all orders should compensate most of the factorization scale dependence.
The resummation of the leading logarithms of this class of NS contributions has been
performed in Ref.~\cite{Presti:2014lqa}. However, these contributions
are not yet implemented in the current version of \texttt{TROLL}.

We now turn to our proposal for the perturbative uncertainty of our resummed results.
We consider $\psi$-soft with AP2 and with the natural choice for the constants, Eq.~\eqref{eq:Cres},
as our best option for threshold resummation. However, the other variants of $\psi$-soft
have the same formal accuracy and allow us to estimate the uncertainty from $1/N$ terms and subleading logarithmic terms.
We therefore suggest to consider, for each of the central scales $\mu_0$ in Eq.~\eqref{eq:centralscales},
the envelope of the canonical 7-point scale variations and the 6 variants of $\psi$-soft resummation (we exclude $N$-soft from the computation).
This corresponds to a total of $7\cdot 6=42$ cross section points,\footnote
{The reader should not be scared by the number of cross section points needed: the code \texttt{TROLL} is very fast
and computes all of them in less than a second.}
from which one takes the highest and the lowest
cross sections as the maximum and minimum of the uncertainty band. As an example, we highlighted in Tab.~\ref{tab:res}
those 42 cross sections entering in the error band computation for the central scale $\mu_0=\mh/2$.
We conventionally take our default best option (shown in red in Tab.~\ref{tab:res}) as the central prediction.

\begin{table}[t]
  \centering
  \begin{tabular}{l|llll}
    & $\mu_0=\mh/4$ & $\mu_0=\mh/2$ & $\mu_0=\mh$ & $\mu_0=2\mh$ \\
    \midrule
             LO & $ 18.6^{+5.8}_{-3.9} $ & $ 16.0^{+4.3}_{-3.1} $ & $ 13.8^{+3.2}_{-2.4} $ & $ 11.9^{+2.5}_{-1.9} $ \\[1ex]
            NLO & $ 44.2^{+12.0}_{-8.5} $ & $ 36.9^{+8.4}_{-6.2} $ & $ 31.6^{+6.3}_{-4.8} $ & $ 27.5^{+4.9}_{-3.9} $ \\[1ex]
           NNLO & $ 50.7^{+3.4}_{-4.6} $ & $ 46.5^{+4.2}_{-4.7} $ & $ 42.4^{+4.6}_{-4.4} $ & $ 38.6^{+4.4}_{-4.0} $ \\[1ex]
        N$^3$LO & $ 48.1^{+0.0}_{-7.5} $ & $ 48.1^{+0.1}_{-1.8} $ & $ 46.5^{+1.6}_{-2.6} $ & $ 44.3^{+2.5}_{-2.9} $ \\[1ex]
\midrule
          LO+LL & $ 24.0^{+8.9}_{-6.8} $ & $ 20.1^{+6.2}_{-5.0} $ & $ 16.9^{+4.5}_{-3.7} $ & $ 14.3^{+3.3}_{-2.8} $ \\[1ex]
        NLO+NLL & $ 46.9^{+15.1}_{-12.6} $ & $ 46.2^{+15.0}_{-13.2} $ & $ 46.7^{+20.8}_{-13.8} $ & $ 47.3^{+26.1}_{-15.8} $ \\[1ex]
      NNLO+NNLL & $ 50.2^{+5.5}_{-5.3} $ & $ 50.1^{+3.0}_{-7.1} $ & $ 51.9^{+9.6}_{-8.9} $ & $ 54.9^{+17.6}_{-11.5} $ \\[1ex]
N$^3$LO+N$^3$LL & $ 47.7^{+1.0}_{-6.8} $ & $ 48.5^{+1.5}_{-1.9} $ & $ 50.1^{+5.9}_{-3.5} $ & $ 52.9^{+13.1}_{-5.3} $ \\[1ex]
  \end{tabular}
  \caption{Fixed-order results and their scale uncertainty together with resummed results and their uncertainty
    (as given by the envelope of prescription and scale variations) for four choices of the central scale,
    for $\mh=125$~GeV at LHC with $\sqrt{s}=13$~TeV. All cross sections are in pb.}
  \label{tab:results}
\end{table}
\begin{figure}[t]
  \centering
  \includegraphics[width=0.495\textwidth,page=2]{images/hxswg_hadr_xsec_res_eff_125_13_PDF4LHC15_uncertainty_summary3.pdf}
  \includegraphics[width=0.495\textwidth,page=1]{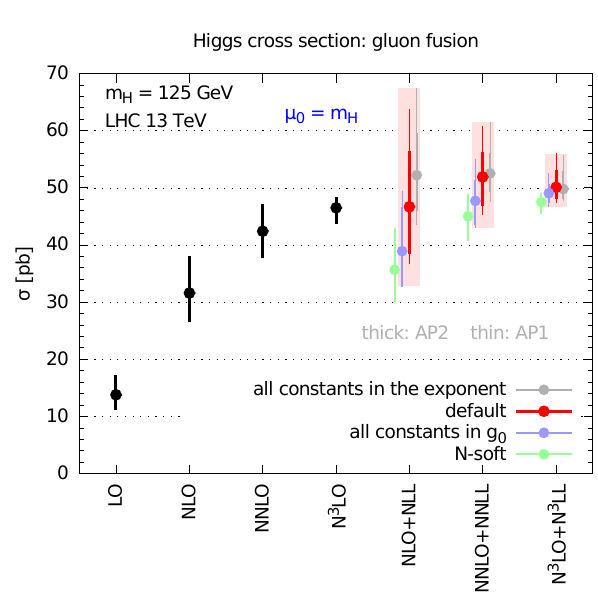}\\
  \includegraphics[width=0.495\textwidth,page=3]{images/hxswg_hadr_xsec_res_eff_125_13_PDF4LHC15_uncertainty_summary3.pdf}
  \includegraphics[width=0.495\textwidth,page=4]{images/hxswg_hadr_xsec_res_eff_125_13_PDF4LHC15_uncertainty_summary3.pdf}
  \caption{Higgs cross section at 13 TeV in the rescaled effective theory (rEFT), for four different choices of the central scale $\muf=\mur$:
    at the top we show $\mh/2$ and $\mh$, while at the bottom  $\mh/4$ and $2\mh$.
    The uncertainty on the fixed-order predictions and on $N$-soft comes solely from scale variation,
    as well as the thick uncertainty on the $\psi$-soft AP2 results. The thinner bands correspond to the 7-point scale variation
    envelope on the $\psi$-soft AP1 instead, whose central value is not shown.
    The light-red rectangles are the envelope of all $\psi$-soft variants,
    corresponding to the 42-point uncertainty described in the text.}
  \label{fig:13TeV}
\end{figure}
We then report in Tab.~\ref{tab:results} the cross section at fixed LO, NLO, NNLO and N$^3$LO accuracy, and its resummed counterpart at
LO+LL, NLO+NLL, NNLO+NNLL and N$^3$LO+N$^3$LL accuracy, for the four central scale choices of Eq.~\eqref{eq:centralscales}.
The error on the fixed-order is computed according to the canonical 7-point variation, while at resummed level
we use our 42-point variation.
The same results are shown as plots in Fig.~\ref{fig:13TeV}.

Let us first comment the fixed-order results. Ignoring the LO which contains too few information for being predictive,
we can investigate the convergence pattern of the fixed-order perturbative expansion when going from NLO to NNLO and to N$^3$LO,
relative to the scale uncertainty.
For ``large'' central scales, $\mu_0=\mh$ and $\mu_0=2\mh$, NNLO is a large correction and its central value is not covered by the NLO uncertainty band. The N$^3$LO is a smaller correction, a sign that the series is converging (at least asymptotically),
but for $\mu_0=2\mh$ its central value is not covered by the NNLO uncertainty band.
For $\mu_0=\mh/2$, the convergence pattern is improved, now with the central NNLO contained in the NLO band,
and the central N$^3$LO contained in the NNLO band.
However, for instance, the central N$^3$LO and its band are \emph{not} contained in the NLO band (they do not even overlap).
At $\mu_0=\mh/4$ the convergence pattern seems further improved, however the N$^3$LO error is very asymmetric
and large (same size of the NNLO error).
Additionally, the N$^3$LO results at the four central scales shown in Table~\ref{tab:results}
are barely compatible (if one had chosen $\mu_0=4\mh$ the result would not be compatible with the one at $\mu_0=\mh/2$).
This analysis shows that the estimate of the uncertainty from missing higher orders using canonical 7-point scale variation is not reliable at fixed order.
This is perhaps not surprising, as scale variation provides a very crude estimate of the uncertainty from missing higher orders,
since it is based on arbitrary variation of a (not necessarily significant) subset of known coefficients.

On the other hand, resummation allows for a different way of estimating the effect of missing higher orders,
which is not purely based on scale variation.
We observe that, for each choice of the central scale $\mu_0$, the uncertainty of the resummed
results from NLO+NLL onwards covers the central value and at least a portion of the band of the next (logarithmic) order.
In fact, with the exception of the choice $\mu_0=\mh/4$ (the pathological behaviour of which seems to be driven by the N$^3$LO contribution),
the NNLO+NNLL band is fully contained in the NLO+NLL band, and the N$^3$LO+N$^3$LL band is fully contained in the NNLO+NNLL band.
We also note a systematic reduction of the scale uncertainty when going from one logarithmic order to the next.

We also observe that the resummed results at each order are all compatible among the different choices of the central scale $\mu_0$,
thereby showing little sensitivity on $\mu_0$.
It is true that at extreme choices of $\mu_0$  the error bands become very asymmetric
and lead to higher values of the cross section at large $\mu_0$ and to lower values of the cross section at small $\mu_0$;
nevertheless, a region of overlap always exists.

We note that our observations on the behaviour of the resummed results would still hold if one considers a less conservative option,
namely our default $\psi$-soft resummation with AP2 and the natural choice for the constants, Eq.~\eqref{eq:Cres},
corresponding to the red dots and the thick red bands in the plots. In fact, with this option the errors would look
more natural, especially at large $\mu_0$ where the AP1 variation (thinner band) increases the size of the error band
dramatically.
It is interesting to observe that the different options for the position of the constants, while giving
a large spread at NLO+NLL, is of little importance at higher orders, especially at N$^3$LO+N$^3$LL.
This is particularly true for smaller $\mu_0$, while at larger $\mu_0$ the version with all constants
in $g_0$, Eq.~\eqref{eq:constantsg0}, gives a slightly smaller cross section.

We finally observe that, in many respects, the choice $\mu_0=\mh/2$ seems optimal,
in full agreement with previous analyses, e.g.~\cite{Anastasiou:2016cez}.
The convergence of the fixed-order is already quite good, and the convergence of the resummed result
is very good. The error band at N$^3$LO+N$^3$LL is smaller than for other central scales,
but compatible with the results computed at different values of $\mu_0$.
On top of these \emph{a posteriori} observations, one could determine \emph{a priori}
an optimal choice for the scale by requiring that the partonic coefficient functions do not contain large logarithms,
such that possible logarithmic enhancements are minimized.
Factorization and renormalization scales typically appear together with threshold logarithms,
in the form (in Mellin space) $\ln(\mu^2 N^2/\mh^2)$. From a saddle point analysis~\cite{Bonvini:2012an},
we know that the Mellin space cross section is dominated by a single value of $N=N_{\rm saddle}$,
leading to an optimal choice for the scales $\mu_0\simeq\mh/N_{\rm saddle}$.
In the present case $N_{\rm saddle}\simeq2$, so we find that the scale that minimizes the size of the logarithms
is close to $\mu_0=\mh/2$.
Similar conclusions have been obtained from $z$-space arguments~\cite{Anastasiou:2002yz}
and within an effective theory framework~\cite{Ahrens:2008nc,Bonvini:2014qga}.
As the process gets closer to threshold, $N_{\rm saddle}$ grows and the optimal scale gets correspondingly smaller:
we have indeed verified that closer to threshold (larger $\mh$ or smaller collider energy)
the perturbative convergence is much improved at smaller scales.

Given that the way of estimating the uncertainty is very conservative, and successful at previous orders,
the uncertainty on the N$^3$LO+N$^3$LL at $\mu_0=\mh/2$ seems reasonably trustful.
To be even more conservative, one can symmetrize the error, so rather than $48.5^{+1.5}_{-1.9}$~pb
we can quote $48.5\pm1.9$~pb as most reliable prediction for the inclusive Higgs cross section (in the rEFT setup).
Note also that at $\mu_0=\mh/2$ the effect of adding the resummation to the N$^3$LO on the central value
is rather small, $+0.4$~pb, corresponding to $+0.8\%$, which is not covered by the fixed-order uncertainty.
However, we find the asymmetric error on the N$^3$LO hard to trust.
Indeed, this is due to the vicinity of a stationary point in the scale dependence,
which in fact shows that such a scale error is not reliable.
Thus, at the very least, one should symmetrize it, leading to $48.1\pm1.8$~pb,
which is then compatible with what we obtain from the resummation procedure.

\begin{figure}[t]
  \centering
  \includegraphics[width=0.495\textwidth,page=2]{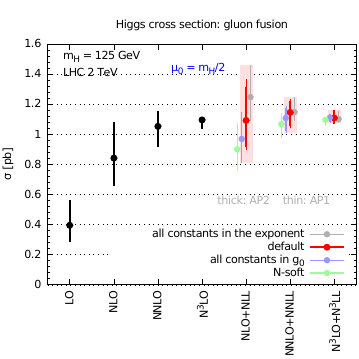}
  \includegraphics[width=0.495\textwidth,page=2]{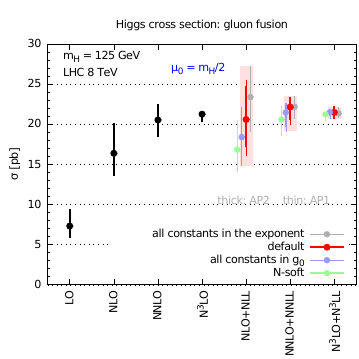}\\
  \includegraphics[width=0.495\textwidth,page=2]{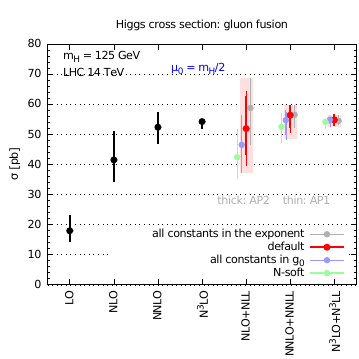}
  \includegraphics[width=0.495\textwidth,page=2]{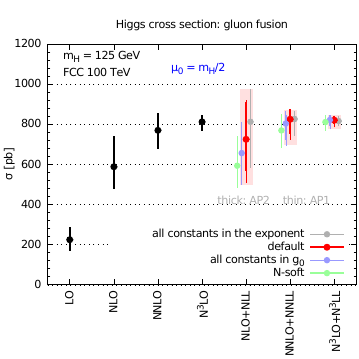}
  \caption{Higgs cross section in the rescaled effective theory (rEFT), for four different values of the collider energy:
    $\sqrt{s}=2,8,14,{\rm and}\,100$~TeV. The central scale is $\muf=\mur=\mh/2$.}
  \label{fig:energy}
\end{figure}
To emphasize the robustness of our method for the estimate of the perturbative uncertainty,
we repeat our analysis for different collider energies: $\sqrt{s} = 2, 8, 14$ and $100$~TeV.
We show in Fig.~\ref{fig:energy} the analogous of the $\mu_0=\mh/2$ plot of Fig.~\ref{fig:13TeV}
for the aforementioned collider energies. We observe the same pattern found at $13$~TeV.
Note that for a FCC-like energy of $100$~TeV the smaller values of $z$ accessible at that energy
will make the correct inclusion of small-$z$ effects (at fixed-order or resummed levels) very important.
We stress in particular that the rEFT is not able to predict the small-$z$ behaviour correctly,
and one should revert to the full theory for an appropriate description.

\section{Other ways to estimate the uncertainty from missing higher orders}\label{sec:missing_HO}

In the previous section we have proposed a robust way of estimating the uncertainty from missing higher orders
using scale variation and variation of subleading terms in the all-order resummation.
In this section we want to explore alternative strategies to estimate the uncertainty from missing higher orders
which do not rely on arbitrary variations of the perturbative ingredients.
In particular, we will consider in Sect.~\ref{sec:CH} the Cacciari-Houdeau method~\cite{Cacciari:2011ze,Forte:2013mda,Bagnaschi:2014wea}
for estimating the theory uncertainty using a Bayesian approach to infer the uncertainty
from the progression of the perturbative expansion.
Then, in Sect.~\ref{sec:CAA}, we will follow the idea of Ref.~\cite{David:2013gaa}
to apply convergence acceleration algorithms to the perturbative series (either the fixed-order or the resummed one)
to estimate the truncation error. 
We shall then compare the findings of these methods with our results of Sect.~\ref{sec:results}.

\subsection{The Cacciari-Houdeau method}
\label{sec:CH}

In Ref.~\cite{Cacciari:2011ze} Cacciari and Houdeau proposed a statistical model
for the interpretation of theory errors, from which one can compute the uncertainty on the
truncated perturbative series for a given degree of belief (DoB) given the first terms in the expansion.
Among the assumptions of the model, there is the fact that all coefficients $c_i$ of the expansion are
bounded by a common value $\bar c$. To account for potential power growth of the coefficients,
the expansion parameter $\as$ is rescaled, giving the expansion
\beq\label{eq:CHser}
\sigma = \sigma_{\rm LO} \sum_{k=0}^\infty c_k(\lambda)\, \(\frac{\as}{\lambda}\)^k.
\eeq
In Eq.~\eqref{eq:CHser} we have factored out the LO cross section $\sigma_0$ such that the sum starts from $k=0$,
with $c_0=1$.
The coefficients $c_k$ depend on the rescaling factor $\lambda$, which should be determined
such that the bound $\bar c$ exists. We come back on the determination of $\lambda$ later in this section.
This is the original method, denoted CH.

It has been noted~\cite{Cacciari:2011ze} that the assumption that all $c_k$ are bounded
is surely broken by the presence of known factorial growths in the coefficients, such as those due to renormalons.
However, it was pointed out that the growth will set in at a high order, so for practical applications
at low orders ignoring it is harmless.
However, in Ref.~\cite{Bagnaschi:2014wea} this factorial growth is instead included in the definition
of the series expansion,
\beq\label{eq:CHmodser}
\sigma = \sigma_{\rm LO} \sum_{k=0}^\infty b_k(\lambda,k_0)\, (k+k_0)!\,\(\frac{\as}{\lambda}\)^k,
\eeq
where the new coefficients $b_k$ do not contain the factorial growth anymore.
This is the modified Cacciari-Houdeau method, denoted $\CHbar$.
We included an explicit offset $k_0$ in the factorial: this is useful since different observables will
have the factorial growth starting at different orders.

Note that in the original works~\cite{Cacciari:2011ze,Bagnaschi:2014wea} there is a distinction
between observables starting at different orders $\as^l$.
In our case, having factored out the LO, the complications arising from this distinction go away.
Factoring out a power of $\as$ does not change the error estimate in the original CH method,
so our Eq.~\eqref{eq:CHser} gives identical results as those obtained with the original formulation of Ref.~\cite{Cacciari:2011ze}.
On the other hand, due to the presence of the factorial, 
factoring out a power of $\as$ in the modified $\CHbar$ method
would make a difference if the factorial is left unchanged.
The inclusion of the offset $k_0$ in the factorial in Eq.~\eqref{eq:CHmodser} accounts for this difference
and allows to exactly reproduce the results of Ref.~\cite{Bagnaschi:2014wea} for the specific choice $k_0=l-1$.

Given the set of N$^k$LO coefficients $c_0,\ldots,c_k$ (or $b_0,\ldots,b_k$ in the modified version)
one can construct the credibility interval $[-d_k^{(p)},d_k^{(p)}]$ in which the remainder of the perturbative series
is expected to lie with DoB $p$. We have~\cite{Cacciari:2011ze,Bagnaschi:2014wea}
\begin{align}
  &\text{CH:} & d_k^{(p)}(\lambda) &= \sigma_{\rm LO}\, \(\frac{\as}{\lambda}\)^{k+1} \max(|c_0|,\ldots,|c_k|) \,F(k,p) \label{eq:CHerr}\\
  &\CHbar: & d_k^{(p)}(\lambda,k_0) &= \sigma_{\rm LO}\, \(\frac{\as}{\lambda}\)^{k+1} \max(|b_0|,\ldots,|b_k|) \,(k+k_0+1)!\, F(k,p) \label{eq:CHbarerr}
\end{align}
with
\beq
F(k,p) =
\begin{cases}
  \frac{k+2}{k+1} \,p &p\leq\frac{k+1}{k+2} \\
  \[(k+2)(1-p)\]^{-1/(k+1)} &p>\frac{k+1}{k+2}
\end{cases}.
\eeq
Note that these simple analytic expressions assume that the remainder of the perturbative series is dominated by the next higher order;
this approximation is good if the expansion parameter $\as/\lambda$ is small enough, but breaks down for small values of $\lambda$.
In such cases, one should use the full result~\cite{Cacciari:2011ze}, which is not expressible in closed analytic form,
or interpret the resulting uncertainty as arising just from the next order.

It now remains to determine the values of $k_0$ and $\lambda$.
Regarding $k_0$, in Ref.~\cite{Bagnaschi:2014wea} it is argued that for a series
starting at order $\as^l$ the proper value is $k_0=l-1$.
This follows from the observation that for weak processes starting at order $\as^0$
the renormalon factorial growth behaves as $\as^k(k-1)!$.
Therefore, for Higgs production which starts at order $\as^2$ the suggested value is $k_0=1$.
However, we point out that what matters here is where the perturbative corrections to gluon propagators
start rather than the power of $\as$ at LO.
Indeed, for processes like Drell-Yan, the gluon appears first at NLO, and the correction to the gluon propagator only at NNLO,
so indeed the factorial growth is delayed by one order and $k_0=-1$.
However, for Higgs production, there are gluons already at LO, so the first correction appears at NLO,
from which one can conclude that $k_0=0$.
We adopt here the latter choice ($k_0=0$) rather than the default option of Ref.~\cite{Bagnaschi:2014wea} ($k_0=1$).

We now turn to the determination of $\lambda$. This parameter is in principle a free parameter,
but it must be such that there exists a bound for the coefficients of the expansion.
In Ref.~\cite{Bagnaschi:2014wea} $\lambda$ is determined from a survey over several processes, giving $\lambda=0.6$.
However, different perturbative expansions can behave in very different ways, and in particular
the perturbative expansion for Higgs production in gluon fusion is badly behaved, so a separate treatment is to be preferred.
Ideally, the value of $\lambda$ should be determined according to the asymptotic behaviour of the perturbative coefficients.
Since this is unfortunately not known, we follow the proposal of Ref.~\cite{Forte:2013mda}, where the value of $\lambda$
is fitted such that the perturbative coefficients (in absolute values) are all of the same size.
In fact, as we also confirmed, it is convenient to exclude the first coefficient from the fit,
on the ground that the LO result is not in line with the next orders (it is much smaller),
and the fact that this fit aims at guessing the asymptotic behaviour of the coefficients.
In the results that follow, we will then use for each method (CH and $\CHbar$) and for each central scale the
value of $\lambda$ obtained by such fit.
As we have observed, however, this determination of $\lambda$ risks being \emph{ad hoc} and therefore may result in a somewhat biased error.

\begin{figure}[t]
  \centering
  \includegraphics[width=\textwidth,page=1]{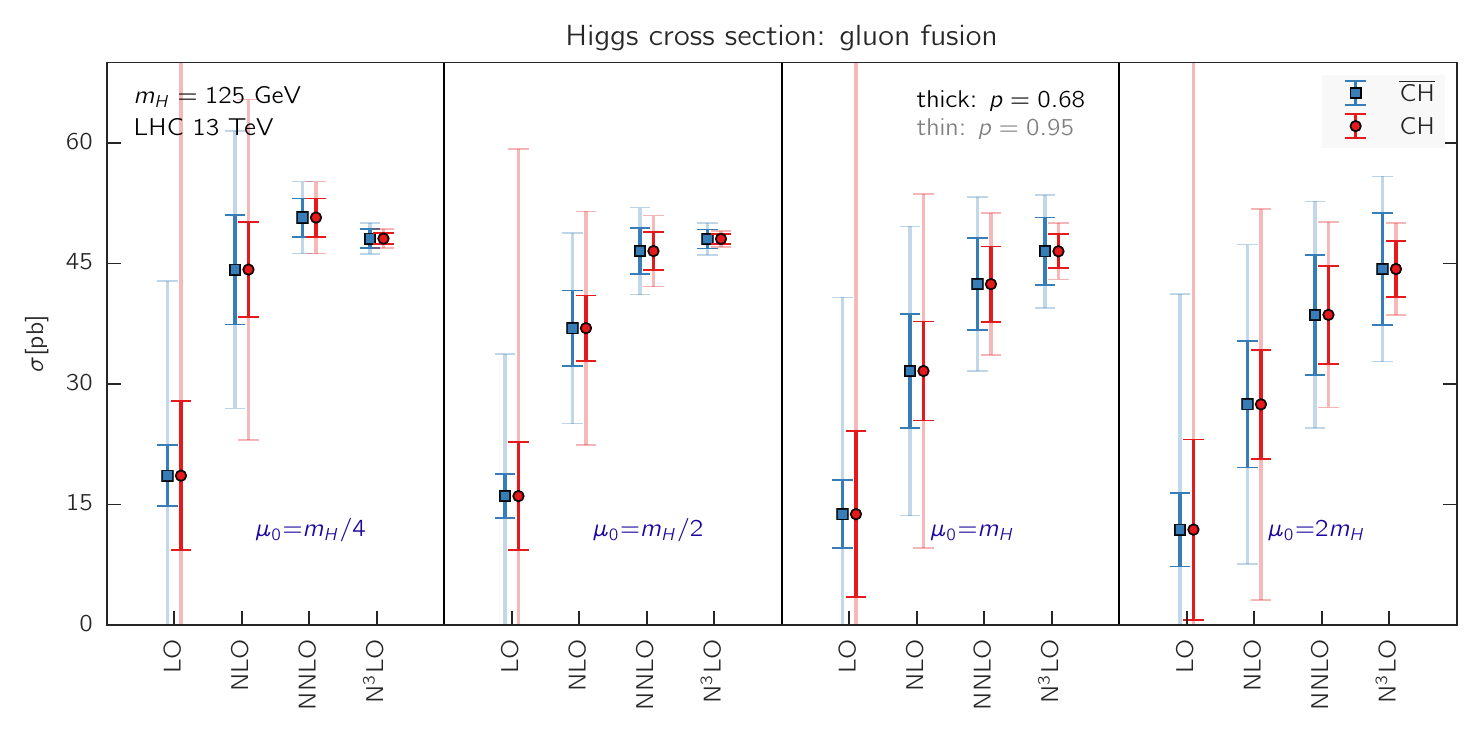}
  \caption{The CH (red) and $\CHbar$ (blue) errors on the LO, NLO, NNLO and N$^3$LO cross sections for the four scales
    $\muf=\mur=\mh/4, \mh/2, \mh, 2\mh$ (from left to right). For the four values of the scales,
    the fitted values of $\lambda$ are respectively $0.44, 0.46, 0.24, 0.17$ for CH
    and $1.08, 1.14, 0.58, 0.41$ for $\CHbar$.
    Thicker bands correspond to $68\%$ DoB, while thinner bands correspond to $95\%$ DoB.}
  \label{fig:CH}
\end{figure}
In Fig.~\ref{fig:CH} we show the four results at LO, NLO, NNLO and N$^3$LO
for the four scales $\muf=\mur=\mh/4, \mh/2, \mh, 2\mh$, each with the two versions (CH and $\CHbar$)
of the Cacciari-Houdeau uncertainty.
We observe that the CH uncertainty is larger than the $\CHbar$ one at LO and NLO, but is smaller at NNLO and N$^3$LO:
this effect originates from the factorial contribution, which changes the relative weight of the individual orders
in the determination of the uncertainty.
In this respect, the $\CHbar$ uncertainty at N$^3$LO is more conservative than the CH one.
We also note that the $68\%$ DoB uncertainty (thicker band) is able to cover the next order only at NNLO, while for lower
orders only the $95\%$ DoB uncertainty (thinner band) works (except at LO for $\CHbar$).
We also see that for small scales $\muf=\mur=\mh/4, \mh/2$ the uncertainty shrinks considerably as the perturbative
order increases, an indication that the series is converging.
For larger scales, $\muf=\mur=\mh, 2\mh$, the observed pattern is much worse and, as a consequence,
the uncertainty band of the N$^3$LO is still large.

\begin{table}[t]
  \centering
  \begin{tabular}{l|llll}
    & $\mu_0=\mh/4$ & $\mu_0=\mh/2$ & $\mu_0=\mh$ & $\mu_0=2\mh$ \\
    \midrule
    CH       & $ 48.1\pm0.7 (1.2) $ & $ 48.1\pm0.6 (1.0) $ & $ 46.5\pm2.1 (3.5) $ & $ 44.3\pm3.5 (5.8) $ \\[1ex]
    $\CHbar$ & $ 48.1\pm1.2 (1.9) $ & $ 48.1\pm1.2 (2.0) $ & $ 46.5\pm4.2 (7.0) $ & $ 44.3\pm6.9 (11.5) $ \\[1ex]
  \end{tabular}
  \caption{N$^3$LO results (in pb) and their CH and $\CHbar$ uncertainties at $68\%$ DoB ($95\%$ DoB in brackets) for $\mh=125$~GeV at LHC with $\sqrt{s}=13$~TeV.}
  \label{tab:CH}
\end{table}
In Tab.~\ref{tab:CH} we report the value of N$^3$LO cross section together with its uncertainty as obtained with CH and $\CHbar$,  for four different choices of the central scale. 
We indicate both $68\%$ and $95\%$ DoBs (the latter in brackets). 
If we focus on the default choice $\mh/2$, we note that the estimate of the uncertainty due
to missing higher orders as obtained combining scale and resummation 
uncertainties (see N$^3$LL+N$^3$LO in Tab.~\ref{tab:results}) is in agreement with the $\CHbar$ uncertainty estimate
at $95\%$ DoB (while CH provides with a smaller uncertainty). In contrast, as previously noted, scale variation on its own
(e.g.\ N$^3$LO in Tab.~\ref{tab:results}) provides us with highly asymmetric error,
which appears to underestimate the upper portion of the uncertainty band.

\subsection{Convergence acceleration algorithms}
\label{sec:CAA}

In this last section we want to explore another method to gain information on the
uncertainty from missing higher orders by estimating the sum of the perturbative series.
Our strategy is based on convergence acceleration algorithms: given a sequence which converges
to some limit there exist several algorithms which transform
the sequence into new ones with possibly faster convergence.
Non-linear sequence transformations usually provide faster convergence than linear transformations.
We follow an idea by David and Passarino~\cite{David:2013gaa} and we apply some of these methods to the sequence of partial sums
of the perturbative expansion of the cross section.

Following Ref.~\cite{Weniger:2003}, we consider the following sequence transformation:
\beq\label{eq:WenigerG}
{\cal G}_k^{(n)}(q_m,s_n,\omega_n) = \frac
{\displaystyle\sum_{j=0}^k(-1)^j\binom{k}{j}\prod_{m=1}^{k-1}\frac{n+j+q_m}{n+k+q_m}\frac{s_{n+j}}{\omega_{n+j}}}
{\displaystyle\sum_{j=0}^k(-1)^j\binom{k}{j}\prod_{m=1}^{k-1}\frac{n+j+q_m}{n+k+q_m}\frac{1}{\omega_{n+j}}},
\eeq
where $s_n=\sum_{i=0}^na_i$ is the $n$-th partial sum, and $a_i$ the series coefficients.
For $k>0$, it provides a non-trivial transformation of the original sequence.
The non-linear transformation Eq.~\eqref{eq:WenigerG} depends on the function $q_m$, and
for particular forms of it reduces to transformations widely studied in the literature~\cite{Weniger:2003}, e.g.\
$q_m=\beta>0$ gives a Levin transformation ${\cal L}_{k}^{(n)}(\beta,s_n,\omega_n)$,
$q_m=\beta+m-1$ gives a Sidi transformation ${\cal S}_{k}^{(n)}(\beta,s_n,\omega_n)$,
$q_m=\beta-m+1$ ($\beta>0$) gives a ${\cal M}_{k}^{(n)}(\beta,s_n,\omega_n)$ transformation,
and $q_m=\beta+(m-1)/\alpha$ gives a transformation ${\cal C}_{k}^{(n)}(\alpha,\beta,s_n,\omega_n)$
which interpolates between $\cal L$ and $\cal S$ depending on the value of $\alpha$.\footnote
{We refer the Reader to Ref.~\cite{Weniger:2003} for an exhaustive explanation of the notation and an extensive list of references.}
These transformations further depend on the function $\omega_n$, which is related to the remainder estimate
of the truncated sequence. In our case we make no assumptions about the asymptotic behaviour of the perturbative
series of the Higgs cross section, mainly because the knowledge of only the first 4 terms is not sufficient
to guarantee that the asymptotic behaviour has set in.
Therefore, we consider various forms of the remainder estimate, leading to a number of variants of the aforementioned methods~\cite{Weniger:2003}:
$u$-type variant given by $\omega_n=(\beta+n)a_n$ (with an overall minus sign for the $\cal M$ transformation),
$t$-type variant given by $\omega_n=a_n$,
$d$-type variant given by $\omega_n=a_{n+1}$,
and $v$-type variant given by $\omega_n=a_na_{n+1}/(a_n-a_{n+1})$.

For $u$- and $t$-type variants, the transformation ${\cal G}_k^{(n)}$ requires the knowledge of $k+n+1$ terms in the sequence,
while for $d$- and $v$-type variants $k+n+2$ terms are needed.
At such low orders, several transformations turn out to be redundant; after removing equivalent transformations,
the non-trivial remaining ones are
$_u{\cal L}_1^{(2)}$,
$_u{\cal L}_2^{(1)}$,
$_u{\cal L}_3^{(0)}$,
$_u{\cal S}_3^{(0)}$,
$_u{\cal M}_3^{(0)}$,
$_u{\cal C}_3^{(0)}$,
$_t{\cal L}_1^{(2)}$,
$_t{\cal L}_2^{(1)}$,
$_t{\cal L}_3^{(0)}$,
$_t{\cal S}_3^{(0)}$,
$_t{\cal M}_3^{(0)}$,
$_t{\cal C}_3^{(0)}$ and
$_d{\cal L}_2^{(0)}$,
where we wrote the variants as a left index.
For comparison, the two methods retained in Ref.~\cite{David:2013gaa} correspond to
$_t{\cal S}_2^{(1)}$ ($={}_t{\cal L}_2^{(1)}={}_t{\cal M}_2^{(1)}={}_t{\cal C}_2^{(1)}$), known as Levin $\tau$ transformation,
and $_d{\cal S}_2^{(0)}$ ($={}_d{\cal L}_2^{(0)}={}_d{\cal M}_2^{(0)}={}_d{\cal C}_2^{(0)}$), known as Weniger $\delta$ transformation.
We recall that each of these variants further depends on the variable $\beta$, and the $\cal C$ transformations also depend on $\alpha$.

In this study, we use all the non-equivalent transformations listed above and scan for various values of $\beta$ and $\alpha$.
Our idea is that, given the unknown higher-order behaviour of the perturbative series, it is
impossible to choose a particular algorithm over the other ones or tune the parameters of the transformations.
Indeed, non-linear sequence transformations are in general not guaranteed to work in all cases,
and privileging a specific algorithm would require the knowledge of the asymptotic behaviour of the series, which we have not.
However, having at hand several different acceleration algorithms, we can judge \emph{a posteriori}
how stable the estimate of the sum of the series is by comparing among the different predictions:
when most results cluster around the same value, we can expect it to be a reliable estimate of the real sum.

As previously mentioned, there are contributions to the perturbative expansion (at parton level) which grow
factorially (due e.g.\ to renormalons). Even though some of the aforementioned sequence transformations
proved to be effective also in case of factorially divergent series, we also consider here a more standard
method based on Borel summation, where the sum of the divergent perturbative series is defined as
\beq\label{eq:Borel}
\sum_{i=0}^\infty a_i = \int_0^\infty dw \, e^{-w} \sum_{i=0}^\infty \frac{a_i}{i!} w^i.
\eeq
When a finite number of coefficients is known, the Borel sum Eq.~\eqref{eq:Borel} becomes an identity.
To all orders, the left-hand side series can diverge due to renormalons, while the sum on the right-hand side
is typically convergent, due to the factorial that cures the growth.
The integral can then be computed: if the integral is finite,\footnote
{In general, the integrand on the right-hand side of Eq.~\eqref{eq:Borel} can have poles on the integration range.
This problem can be solved by modifying the integration contour and avoiding the singularity from above or below,
leading to a well-known renormalon ambiguity. In our case, the ambiguity is always very small, so we ignore it.}
then the right-hand side of Eq.~\eqref{eq:Borel} can be used to \emph{define} the (Borel) sum of the divergent series.
In our case, we can benefit from the Borel summation by applying a convergence acceleration algorithm to
compute the all-order sum on the right-hand side, thereby curing possible divergences of the original series.
Hence, for each of the methods listed above, we can produce a Borel variant, doubling the available options.

\begin{figure}[t]
  \centering
  \includegraphics[width=0.495\textwidth,page=2]{images/HistoFO_RES_paper.pdf}
  \includegraphics[width=0.495\textwidth,page=3]{images/HistoFO_RES_paper.pdf}\\
  \includegraphics[width=0.495\textwidth,page=1]{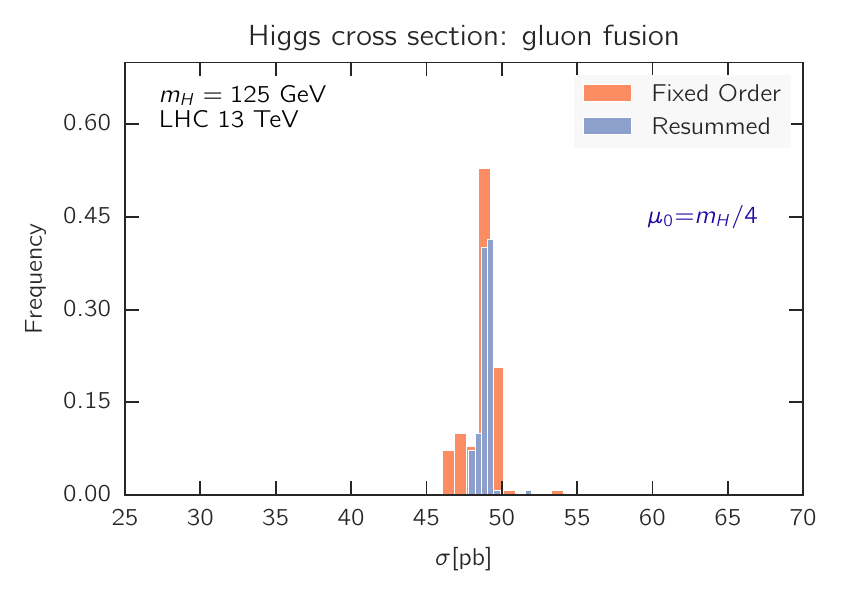}
  \includegraphics[width=0.495\textwidth,page=4]{images/HistoFO_RES_paper.pdf}
  \caption{Distributions of the Higgs cross section at 13 TeV as obtained using the various
    convergence acceleration algorithms described in the text.
    Both the fixed-order (orange) and the resummed (blue) expansions are shown,
    for the four scales $\mur=\muf=\mh/2$ (top left), $\mh$ (top right), $\mh/4$ (bottom left) and $2\mh$ (bottom right).}
  \label{fig:CAA}
\end{figure}

Having implemented all these variants, we apply them to the fixed-order and resummed\footnote
{For the resummation we use $\psi$-soft  AP2, with default option for the constants.}
expansions for each of the previously considered central scales $\mu_0=\mh/4, \mh/2, \mh, 2\mh$.
For each method, we compute the results for several values of $\beta$; applying these methods
to a number of known series we identified $\beta=0.01, 1, 2, 5$ as a sensible set.
For the $\cal C$ transformations we also scan over $\alpha = 0.5, 2, 10, 100$:
this is motivated by the fact that a value of $\alpha$ between $1$ and $+\infty$ interpolates
between the $\cal S$ and $\cal L$ transformations, and we also considered a value outside this range.
Note that $_u{\cal L}_1^{(2)}$ and $_t{\cal L}_2^{(1)}$ do not depend on $\beta$, so these are counted only once.
We collect the results for the estimated sum of the fixed-order and resummed series in the form of histograms in Fig.~\ref{fig:CAA}.

We immediately observe that the distribution of results is narrower for resummed
results than for fixed-order ones. This can be expected since the resummed expansion
is much better behaved than the fixed-order one, as commented extensively in Sect.~\ref{sec:results}.
Additionally, we observe that the distributions for $\mu_0=\mh/4$ and $\mu_0=\mh/2$
are narrower than those at higher scales, a fact which is especially true for the fixed-order expansion.
This confirms the faster convergence of the perturbative series at $\mu_0=\mh/2$ (and also $\mu_0=\mh/4$)
observed in previous sections and discussed in the literature.
The fact that for $\mu_0=\mh$ and $\mu_0=2\mh$ the spread of the results is rather wide
shows that the algorithms considered here are sufficiently different among each other. 
This is important because it validates the set of algorithms we have chosen, which in turn
validates the rather precise results obtained at lower scales.

\begin{table}[t]
  \centering
  \begin{tabular}{l|llll}
    & $\mu_0=\mh/4$ & $\mu_0=\mh/2$ & $\mu_0=\mh$ & $\mu_0=2\mh$ \\
    \midrule
    Fixed-order expansion & $ 48.7\pm1.0 $ & $ 48.7\pm1.2 $ & $ 46.3\pm4.6 $ & $ 44.6\pm9.3 $ \\[1ex]
    Resummed    expansion & $ 48.9\pm0.5 $ & $ 48.9\pm0.6 $ & $ 50.2\pm1.0 $ & $ 52.6\pm1.6 $ \\[1ex]
  \end{tabular}
  \caption{Mean and standard deviation of the estimates of the all-order sum (in pb) of
    the fixed-order (first row) and resummed (second row) expansions,
    based on the set of convergence acceleration algorithms described in the text,
    for $\mh=125$~GeV at LHC with $\sqrt{s}=13$~TeV.}
  \label{tab:CAA}
\end{table}
To make the discussion more quantitative, we report in Tab.~\ref{tab:CAA} the mean and standard deviation
for each histogram of Fig.~\ref{fig:CAA}.
We warn the Reader that the statistical interpretation of these results is not solid:
in particular, the standard deviations are likely to depend on the set of convergence acceleration algorithms considered.

All the numbers in Tab.~\ref{tab:CAA} come from estimates of the all-order sum of the series,
which should be then the same for all scales and for both the fixed-order and the resummed expansions.
They are indeed all compatible within the quoted errors, except the resummed result at $\mu_0=2\mh$
which is higher than most of the other results: this is just a consequence of the limited statistical meaning
of the error estimates, which does not take into account the shape of the distribution of the results,
which is rather asymmetric in this case.
The smaller standard deviation on the resummed results shows once again that the resummed series converges faster,
as well as the smaller standard deviation on the results at lower scales indicates that using $\mu_0=\mh/2$ or $\mu_0=\mh/4$
leads to a faster convergence, in agreement with the findings of our previous sections.
It is interesting to observe that at both scales $\mu_0=\mh/2$ or $\mu_0=\mh/4$ and for both fixed-order and resummed expansions
the estimate of the sum is basically the same ($48.7$ from fixed-order and $48.9$ from resummation), both
with a small standard deviation (of the order of $1$~pb).
Keeping in mind the limitation of this analysis, we are tempted to consider this result
as a good candidate for the all-order sum of the series.
Interestingly, this result is perfectly compatible with (and very close to) our best N$^3$LO+N$^3$LL
result at $\mu_0=\mh/2$ well within its $\pm1.9$~pb uncertainty.
This provides another valuable validation of our proposal for estimating missing higher-order uncertainty from resummation.
On the other hand, it is not compatible with the N$^3$LO result within its asymmetric
scale-variation band, while it is considering a $\CHbar$ error already at $68\%$ DoB (see Tab.~\ref{tab:CH}).

\section{Conclusions}

We have presented threshold-resummed results for the inclusive Higgs cross section in gluon fusion at N$^3$LL,
matched to an implementation of the recent N$^3$LO result~\cite{Anastasiou:2014vaa,Anastasiou:2014lda,Anastasiou:2015ema,Anastasiou:2016cez}.
We have considered several variants of the resummation as a portal to carefully estimate subleading effects at higher orders.
We have proposed a conservative estimate of the uncertainty from missing higher orders based on the envelope of the resummed predictions
obtained using the various resummation variants, as well as canonical scale-variation.
We have demonstrated that resummed results with this conservative error manifest a good perturbative convergence,
as opposed to the fixed-order expansion, the convergence of which is very poor relative to the uncertainty coming from a canonical 7-point scale variation.

Despite the conservativeness of our method, we find that the Higgs cross section at 13 TeV, for the central scale $\mur=\muf=\mh/2$, has a small
(yet reliable) uncertainty of $\pm1.9$~pb, which corresponds to $\pm4\%$.
We stress that all choices of central scales considered in this work ($\mh/4$, $\mh/2$, $\mh$ and $2\mh$)
yield results which are compatible within the quoted uncertainty.
The shift in the central value and the uncertainty, though computed within the framework
of the (rescaled) large-$\mt$ effective theory, are likely to remain unchanged after inclusion of quark mass effects
and Electro-Weak corrections.
For the most reliable predictions the inclusion of quark mass effects is important, and can be performed
straightforwardly at resummed level~\cite{Bonvini:2014joa,Schmidt:2015cea} with \texttt{TROLL}.
Moreover, a fully consistent resummed result would require the use threshold-improved parton distribution functions,
which have recently become available~\cite{Bonvini:2015ira}.

We have compared our proposal with different methods for estimating the uncertainty from missing higher orders.
Our findings are summarized in the following table, which refers to the central scale $\mur=\muf=\mh/2$:
\begin{center}
  \begin{tabular}{l|ll}
    order & $\sigma$ [pb] & \\
    \midrule
    N$^3$LO & $48.1^{+0.1}_{-1.8}$ & scale variation \\
    N$^3$LO & $48.1\pm 2.0$ & $\CHbar$ at $95\%$ DoB \\
    N$^3$LO+N$^3$LL & $48.5\pm 1.9$ & scale+resummation variations \\
    all-order estimate & $48.7$ & from accelerated fixed-order series \\
    all-order estimate & $48.9$ & from accelerated resummed series \\
  \end{tabular}
\end{center}
First, we have considered the Cacciari-Houdeau Bayesian approach, which employs
the known perturbative orders to construct a probability distribution for the subsequent unknown order.
In its modified incarnation ($\CHbar$),
the method gives an uncertainty of $\pm2$~pb at $95\%$ degree of belief, fully compatible with the estimate obtained from resummation,
and similar to the fixed-order scale variation uncertainty if the latter is symmetrized.
Second, we have considered several algorithms to accelerate the convergence of the perturbative series,
based on non-linear sequence transformations. 
By performing a survey of different algorithms,
we found that both the fixed-order and resummed series exhibit good convergence properties at $\mh/2$ (and also at $\mh/4$).
Noticeably, the mean of each distribution is very close to the N$^3$LO+N$^3$LL prediction.
In conclusion, these tests provide a solid support to our method, and let us conclude to a high degree of belief
that the all-order Higgs cross section in the rEFT lies within the quoted uncertainty of our N$^3$LO+N$^3$LL result.

\acknowledgments{
We are grateful to Stefano Forte for many useful discussions and feedback. 
We also acknowledge discussions with Richard Ball, Matteo Cacciari,  Alberto Guffanti, Achilleas Lazopoulos, Giovanni Ridolfi and Juan Rojo.
The work of MB and LR is supported by an European Research Council Starting Grant `PDF4BSM'.
The work of SM is partly supported by the U.S. National Science Foundation, under grant PHY-0969510, the LHC Theory Initiative.
}

\appendix
\section{Details on numerical implementations}

\subsection{N$^3$LO implementation in \texttt{ggHiggs}}
\label{app:ggHiggs}

We report in this appendix some technical details on the implementation of the recent N$^3$LO result~\cite{Anastasiou:2016cez}
in the public code \texttt{ggHiggs}~\cite{ggHiggs}.
In the EFT, the cross section factorizes as in Eq.~\eqref{eq:EFT}.
The new N$^3$LO contribution published in Ref.~\cite{Anastasiou:2016cez} is the third order coefficient $\tilde C^{(3)}_{ij}$.
We have coded in \texttt{ggHiggs} the coefficient $\tilde C^{(3)}_{ij}$ for $\muf=\mur=\mh$ as given above.

We observed in Sect.~\ref{sec:n3lo} that the leading small-$z$ logarithm is known from high energy resummation.
This term can be added to the soft expansion, after subtracting the appropriate double counting.
For the $gg$ channel we have explicitly
\beq\label{eq:addln5}
\[\tilde C^{(3)}_{gg}(z)\]_{\text{soft-exp}+\ln^5z} = \[\tilde C^{(3)}_{gg}(z)\]_{\text{soft-exp}}
+ \frac{(2C_A)^3}{5!} \[\frac1z\ln^5\frac1z - \sum_{k=0}^{37}a_k(1-z)^k\],
\eeq
where $a_k$ are the expansion coefficients of $\frac1z\ln^5\frac1z$ in powers of $1-z$, and we assumed that
all the known terms in the soft expansion of $\tilde C^{(3)}_{gg}(z)$ are used.
The last sum in Eq.~\eqref{eq:addln5} removes the doubly counted contributions.
Analogously, the leading small-$z$ logarithm in the full theory, $\frac1z\ln^2z$, can be added to the
soft expansion of the full coefficient $C^{(3)}_{ij}(z)$ subtracting the appropriate double counting.

The requirement of scale invariance enables us to 
recover the scale-dependent  logarithmic terms to be added to the result at central scale.
We compute the scale dependent contributions in $N$ space, thus we need the DGLAP anomalous dimensions
up to NNLO~\cite{Vogt:2004mw,Moch:2004pa} as well as the Mellin moments of the NLO and NNLO coefficient functions.
While the exact expression for the NLO coefficient (in the EFT) is easy to compute,
a formulation of the NNLO Mellin-space coefficient which is valid in the whole complex $N$ plane
is much more challenging~\cite{Blumlein:2005im}.
In the latest version of \texttt{ggHiggs}, v3.1, we use an implementation in terms of Harmonic sums up to weight four
whose analytic continuation for non integer values has been implemented in \texttt{C++} following the method of Refs.~\cite{Blumlein:1998if,Blumlein:2000hw,Blumlein:2005im}.

Note that in a recent version of \texttt{ggHiggs}, v3.0, the scale-dependent contributions were computed as a soft expansion matched to
the exact small-$z$ behaviour. We do not report the details of this (now obsolete) implementation,
but we stress that it was very accurate, with an error on the cross sections of Tab.~\ref{tab:FO}
always less than $0.03$~pb.
The numbers obtained with \texttt{ggHiggs} v3.1 are in good agreement
(within the numerical errors) with those of Ref.~\cite{Anastasiou:2016cez}.

\subsection{Matching and scale dependence of the resummed result}
\label{app:TROLL}

In this section we show how we tested the correct implementation of the matching to fixed order of our resummed result.
To perform a correct matching it is necessary to subtract from the resummed result its fixed-order expansion, Eq.~\eqref{eq:DeltaC}.
It is therefore important to verify that the fixed-order expansion, which has been computed analytically, is indeed correct.
To do so, we compared the analytic fixed-order expansion of the resummation to a numerical expansion of the resummation itself.
We report here the results of this comparison.

\begin{figure}[t]
  \centering
  \includegraphics[width=0.495\textwidth,page=1]{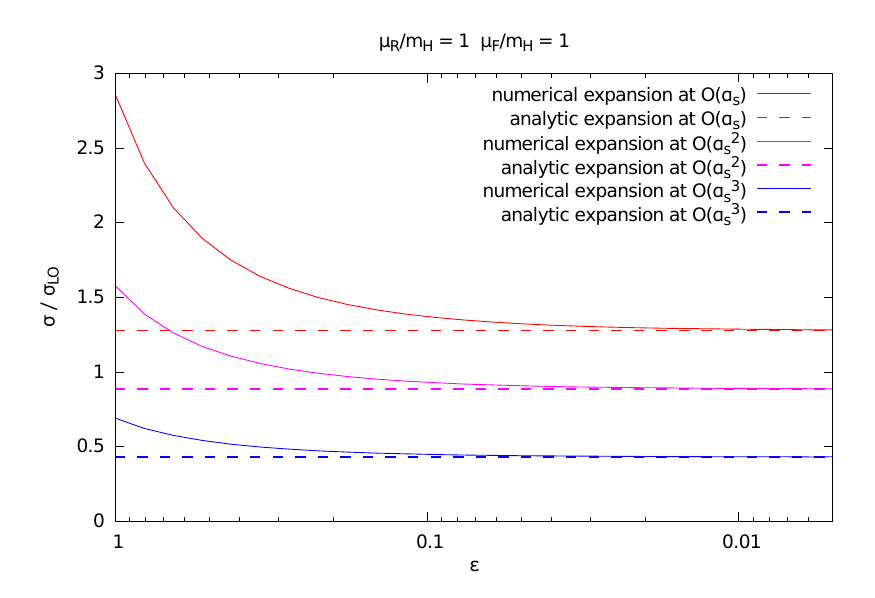}
  \includegraphics[width=0.495\textwidth,page=2]{images/test_scale_dependence.pdf}\\
  \includegraphics[width=0.495\textwidth,page=3]{images/test_scale_dependence.pdf}
  \includegraphics[width=0.495\textwidth,page=4]{images/test_scale_dependence.pdf}
  \caption{Contributions to the cross section (normalised to the LO) from the analytic expansion of the resummation (default $\psi$-soft at N$^3$LL)
    to orders $\as$ (red), $\as^2$ (purple) and $\as^3$ (blue) (dashed lines).
    The numerical expansion, as a function of $\epsilon$, Eq.~\eqref{eq:epsilon}, is also shown as solid lines.}
  \label{fig:test_scales}
\end{figure}

Recalling Eq.~\eqref{eq:DeltaC} we have that the resummed expression can be expanded in powers of $\as$ as
\beq
C_{\rm res}(N,\as) = 1 + \sum_{i=1}^\infty \as^i\, C_{\rm res}^{(i)}(N).
\eeq
For matching to N$^3$LO the expansion coefficients $C_{\rm res}^{(1)}(N)$, $C_{\rm res}^{(2)}(N)$ and $C_{\rm res}^{(3)}(N)$ are needed,
and they have been computed analytically and coded in \texttt{TROLL}.
Each coefficient can be also extracted numerically according to the formula
\beq\label{eq:epsilon}
\as^k\, C_{\rm res}^{(k)}(N) = \lim_{\epsilon\to0} \[\epsilon^{-k}\(C_{\rm res}(N,\as \epsilon) - \sum_{i=0}^{k-1} \(\as\epsilon\)^i\, C_{\rm res}^{(i)}(N)\)\].
\eeq
The expression in squared brackets is of $\Ord(\as^k\epsilon^0)$ and contains corrections which starts at $\Ord(\as^{k+1}\epsilon)$:
therefore, the $\epsilon\to0$ limit suppresses higher-order corrections and isolates the $\Ord(\as^k)$ term.
In Fig.~\ref{fig:test_scales} we show the analytic results for the cross section contributions corresponding to
$\as^k\, C_{\rm res}^{(k)}(N)$ for $k=1,2,3$ together with the numerical expression Eq.~\eqref{eq:epsilon}
as a function of $\epsilon$.
We see that at small $\epsilon$ the numerical expansion reproduces the analytic result
for several non-trivial combinations of the scales $\mur$ and $\muf$.
This represents a strong check of the implementation of the matching to fixed-order for any $\muf$ and $\mur$.
In particular, it ensures that the contribution from the resummation is always one order higher than the fixed order we are matching to.

In order to cross check the scale dependence of the resummed result, we now verify
the explicit scale dependence of the expansion of the resummation to fixed order, which, as we have just verified, is consistent
with the all-order expression.
To do so, we compare the results obtained from the internal implementation of the scale dependence with
an external implementation.
To be precise, we compare the contributions to the cross section
\beq\label{eq:test1}
\sigma^{(k)} \equiv \as^{k+2}(\mur^2) \int_{c-i\infty}^{c+i\infty} \frac{dN}{2\pi i}\,\tau^{-N}\,
\Lum_{gg}(N,\muf^2)\, C_{\text{res}}^{(k)}(N,\muf^2,\mur^2)
\eeq
(where for simplicity we removed overall factors from the definition of $\sigma^{(k)}$) to the analogous computed with $\muf=\mur=\mh$
plus scale dependent terms
\beq\label{eq:test2}
\sigma^{(k)} \equiv \as^{k+2}(\mur^2) \int_{c-i\infty}^{c+i\infty} \frac{dN}{2\pi i}\,\tau^{-N}\,
\Lum_{gg}(N,\muf^2)\, C_{\text{res}}^{(k)}(N,\mh^2,\mh^2) + \Delta\sigma_{\text{scale}}^{(k)}\(\muf^2,\mur^2\).
\eeq
In this way, in the second expression Eq.~\eqref{eq:test2} all the explicit logs of $\mur$ and $\muf$ in the partonic coefficients are set to zero,
and the scale dependent terms are provided with the additional contribution $\Delta\sigma_{\text{scale}}^{(k)}$
which we now construct and which provides a strong cross check of the scale dependence as derived from the resummed expression.
To find $\Delta\sigma_{\text{scale}}^{(k)}$ we impose scale independence of the cross section up to $\Ord(\as^3)$ in the soft limit.
Defining the partonic coefficient $\Delta C_{\text{scale}}^{(k)}$ such that
\beq
\Delta\sigma_{\text{scale}}^{(k)}\(\muf^2,\mur^2\) = 
\as^{k+2}(\mur^2) \int_{c-i\infty}^{c+i\infty} \frac{dN}{2\pi i}\,\tau^{-N}\,
\Lum_{gg}(N,\muf^2)\, \Delta C_{\text{scale}}^{(k)}\(N,\muf^2,\mur^2\)
\eeq
we have
\begin{subequations}\label{eq:scaledepterms}
\begin{align}
  \Delta C_{\text{scale}}^{(1)}\(N,\muf^2,\mur^2\)
  &= 2\gamma_{gg}^{(0)}\ell_F + 2\beta_0 \ell_R\\
  \Delta C_{\text{scale}}^{(2)}\(N,\muf^2,\mur^2\)
  &= 2\(\gamma_{gg}^{(1)} + \gamma_{gg}^{(0)} C_{\rm res}^{(1)} \) \ell_F
    + \(2\beta_1 + 3\beta_0 C_{\rm res}^{(1)} \) \ell_R \nonumber\\
  &+ \gamma_{gg}^{(0)}\(2\gamma_{gg}^{(0)} +\beta_0\) \ell_F^2 + 3\beta_0^2\ell_R^2 +6\beta_0\gamma_{gg}^{(0)}\ell_F\ell_R \\
  \Delta C_{\text{scale}}^{(3)}\(N,\muf^2,\mur^2\)
  &= 2\(\gamma_{gg}^{(2)}+ \gamma_{gg}^{(1)} C_{\rm res}^{(1)} + \gamma_{gg}^{(0)} C_{\rm res}^{(2)}\) \ell_F
    + \(2\beta_2 + 3\beta_1 C_{\rm res}^{(1)} + 4\beta_0 C_{\rm res}^{(2)} \) \ell_R \nonumber\\
  & + \[2\gamma_{gg}^{(1)} \(\gamma_{gg}^{(0)}+\beta_0\)
    + \gamma_{gg}^{(0)} \(2\gamma_{gg}^{(1)} + 2\gamma_{gg}^{(0)} C_{\rm res}^{(1)}+C_{\rm res}^{(1)}\beta_0+\beta_1\)\]\ell_F^2 \nonumber\\
  & + \[7\beta_0\beta_1 + 6\beta_0^2 C_{\rm res}^{(1)}\]\ell_R^2
    + \[6\beta_1\gamma_{gg}^{(0)}+8\beta_0\(\gamma_{gg}^{(1)} + \gamma_{gg}^{(0)} C_{\rm res}^{(1)}\)\]\ell_F\ell_R \nonumber\\
  &+ \frac23 \gamma_{gg}^{(0)} \[\gamma_{gg}^{(0)}\(2\gamma_{gg}^{(0)} +\beta_0\) +2\gamma_{gg}^{(0)}\beta_0 +\beta_0^2\] \ell_F^3 +4\beta_0^3\ell_R^3 \nonumber\\
  &+ 4\beta_0\gamma_{gg}^{(0)}\(2\gamma_{gg}^{(0)} +\beta_0\) \ell_F^2\ell_R
    +12\beta_0^2\gamma_{gg}^{(0)} \ell_F\ell_R^2,
\end{align}
\end{subequations}
 where $C_{\rm res}^{(1)}$ and $C_{\rm res}^{(2)}$ are computed at $\mur=\muf=\mh$,
\beq
\ell_F = \ln\frac{\mh^2}{\muf^2}, \qquad
\ell_R = \ln\frac{\mur^2}{\mh^2},
\eeq
and $\gamma_{gg}^{(k)}$ are the $\Ord(\as^{k+1})$ anomalous dimensions in the soft limit (explicit expressions are given in Ref.~\cite{Vogt:2004mw}).
We have verified that the two expressions Eq.~\eqref{eq:test1} and Eq.~\eqref{eq:test2} give identical results for
any combination of $\mur$ and $\muf$ at orders $k=1,2,3$.
This represents an independent cross-check of the scale dependence of the resummed expression.

\phantomsection
\addcontentsline{toc}{section}{References}
\bibliographystyle{JHEP}
\bibliography{biblio}
\end{document}